\documentclass{iopart}

\usepackage{fleqn}

\usepackage{ifthen}
\usepackage{ifpdf}

\usepackage{latexsym}
\usepackage{amssymb}
\usepackage{bm}

\ifpdf
\usepackage{graphicx}
\usepackage{epstopdf}
\else
\usepackage{graphicx}
\usepackage{epsfig}
\fi

\newcommand{\sinc}{\mbox{sinc}}
\newcommand{\const}{\mbox{const}}
\newcommand{\trc}{\mbox{trace}}

\newcommand{\eexp}{\mbox{e}^}
\newcommand{\bra}{\left\langle}
\newcommand{\ket}{\right\rangle}
\newcommand{\mass}{\mathsf{m}}

\newcommand{\tbox}[1]{\mbox{\tiny #1}}

\newcommand{\pd}[2]{\frac{\partial #1}{\partial #2}}

\newcommand{\be}[1]{\begin{eqnarray}\ifthenelse{#1=-1}{\nonumber}{\ifthenelse{#1=0}{}{\label{e#1}}}}
\newcommand{\ee}{\end{eqnarray}}

\newcommand{\hide}[1]{}

\newcommand{\mpg}[2][\hsize]{\begin{minipage}[b]{#1}{#2}\end{minipage}}
\newcommand{\putgraph}[2][width=\hsize]{\includegraphics[#1]{#2}}

\begin{document}


\hide{

LIST OF FIGURES

--Fig1
step_pic
step_phase

--Fig2
well_pic
well_phase

--Fig3
ring_pic
ring_phase

--Fig4
well_qm_levels
ring_qm_levels

--Fig5
ring_scales

--Fig6
bloch_sites

--Fig7
step_cl_plot
step_qm_plot

--Fig8
well_1-3
well_3-3

--Fig9
well_31-2
well_32-2

--Fig10
ring_3-3
bloch_plot

--Fig11
bloch_1
bloch_2
bloch_3
bloch_4

}


\title[Diffractive energy spreading]
{Diffractive energy spreading \\ and its semiclassical limit}

\author{Alexander Stotland and Doron Cohen}

\address{
Department of Physics, Ben-Gurion University, Beer-Sheva 84005, Israel
}


\begin{abstract}
We consider driven systems where the driving induces jumps
in energy space: (1)~particles pulsed by a step
potential; (2)~particles in a box with a moving wall;
(3)~particles in a ring driven by an electro-motive-force.
In all these cases the route towards quantum-classical correspondence
is highly non-trivial. Some insight is gained by observing that
the dynamics in energy space, where $n$ is the level index,
is essentially the same as that of Bloch electrons
in a tight binding model, where $n$ is the site index.
The mean level spacing is like a constant electric field
and the driving induces long range hopping ${\propto 1/(n-m)}$ .
\end{abstract}

\section{Introduction}

Consider a system which is described by a Hamiltonian $\mathcal{H}(X(t))$,
where the parameter $X(t)$ is time dependent.
For such system the energy $E$ is not a constant of the motion.
Rather, the driving induces spreading in energy space. Assuming that the
system is prepared at $t=0$ in a microcanonical state, one wonders
how the energy distribution $\rho_t(E)$ looks like at a later time.
In particular, one may wonder whether the quantum $\rho_t(E)$
is similar to the corresponding classical distribution.
In the ``quantum chaos" literature it is customary to distinguish
between a classical time scale $\tau_{cl}$ and a quantum breaktime~$t^*$.
The latter goes to infinity in the ``$\hbar\rightarrow0$" limit.
A prototype model is the``quantum kicked rotator" \cite{qkr} where
the energy spreading is diffusive up to $t^*$ while for larger times
one observes saturation due to a dynamical localization effect.

In this work we analyze much simpler systems where the breaktime $t^*$,
if exists, is much larger than any physically relevant time scale.
In fact one may assume that the time~$t$ of the evolution
is comparable with the classical (short) time scale.
In such circumstances one naively would expect
quantum to classical correspondence (QCC).
But in fact the theory is much more complicated \cite{lrt}.
One has to distinguish between
\begin{itemize}
\item Detailed QCC
\item Restricted QCC
\end{itemize}
Detailed QCC means that all the moments ${r=1,2,3,...}$
of the quantum mechanical distribution $\rho_t(E)$ are similar
to the classical result, while restricted QCC refers only 
to the ${r=1,2}$ moments. It turns out that the latter 
are very robust, while the higher moments (${r>2}$) might be
much larger in the quantum case.
Our first challenge would be to find and
to analyze the {\em worst case} for QCC, for which all
the ${r>2}$ moments are classically finite but quantum mechanically
divergent. We would like to see whether in such
circumstances restricted QCC for ${r=1,2}$ survives.

\hide{
Given that the system is prepared in a stationary state at $t=0$,
one can prove that ${\langle\mathcal{H}(t)^2\rangle_0-\langle\mathcal{H}(0)^2\rangle_0
= \langle(\mathcal{H}(t)-\mathcal{H}(0))^2\rangle_0 }$, where $\mathcal{H}(t)$
is the Hamiltonian $\mathcal{H}(X(t))$ is the {\em Heisenberg picture}.
Such relation cannot not be generalized to higher moments because of lack of
commutativity. It should be clear that  ${ \langle(\mathcal{H}(t)-\mathcal{H}(0))^r\rangle_0 }$
is a quantity for which there is a robust QCC.
Using $d\mathcal{H}/dt=\dot{X}V$ where $V\equiv \partial \mathcal{H} / \partial X$
it can be expressed using correlation functions of the perturbation~$V$.}

For completeness of this Introduction we summarize in Appendix~A 
the reason for the robustness of restricted QCC. 
Our interest in QCC is motivated by the wish to develop a better
understanding of driven systems. We would like to explore
examples where QCC is far from obvious even for short times.
In what follows we address 4~problems that in first sight look unrelated:
\begin{itemize}
\item[\bf(1)] Particles that are pulsed by a step potential (Fig.1)
\item[\bf(2)] Particles in a box with a moving wall (Fig.2)
\item[\bf(3)] Particles in an electro-motive-force (EMF) driven ring (Fig.3)
\item[\bf(4)] Wavepacket dynamics of Bloch electrons in a constant electric field
\end{itemize}
In fact we are going to see that problems (1)-(3) share
a common feature: In the classical description
the energy absorption is associated with abrupt {\em jumps}
in phase space. These jumps are reflected in the
quantum dynamics as a strong {\em diffraction} effect.
This diffraction, which takes place in energy space,
is the worst case for Bohr's QCC.
It turns our that problem (1) can be solved exactly,
while problems (2) and (3) reduce essentially to problem (4).
Namely, the dynamics in energy space, where $n$ is the level index,
is essentially the same as that of Bloch electrons
in a tight binding model, where $n$ is the site index.
The mean level spacing is like a constant electric field
and the driving induces long range hopping ${\propto 1/(n-m)}$ .
This tight binding problem has an exact solution.
The objectives of the present work are
\begin{itemize}
\item To highlight the route towards QCC in the case of diffractive energy spreading.
\item To provide solutions and numerical demonstrations to the prototype problems.
\item To shed new light of the EMF-driven ring problem.
\item To illuminate the limitations of linear response theory in the mesoscopic context.
\end{itemize}
The paper is structured accordingly.

A few words are in order regarding the literature.
The quantum treatment of the ``moving wall" problem
has started with Refs.\cite{doescher,makovski},
that were aimed in finding the steady state solutions
for an expanding well. The interest in this model
has further evolved within the study of
the Fermi acceleration problem \cite{jose}
where the wall is oscillating.
Recently the non-trivial features of the
parametric \cite{prm} and of the time-dependent
wavepacket dynamics \cite{wld} were illuminated.
In the latter publication a satisfactory mathematical
treatment of the non-stationary dynamics has not been introduced.
Also the problem of Bloch electrons in a constant
electric field has a long history. The concept
of a Stark ladder was introduced by Wannier \cite{wannier}
to describe the energy spectrum of a periodic system
in an electric field. Since that time it has become
the subject of controversy
\cite{zak-1968,zak-1968-168,rabinovitch-zak-1972,rabinovitch-1977,
emin-hart-1987,hart-1988,hart-emin-1988,Mendez-1988,zak-1991,
leo-1991,zak-1996}.
Eventually it has been realized that
the electric field localizes the motion of the electrons,
and induces a periodic oscillatory motion.

\section{Energy jumps in phase space}


If a Gaussian wavepacket is moving in a smooth potential,
then its Wigner function evolves in a smooth manner
which favors detailed QCC. But we would like to consider
the ``worst case" for QCC. Let us assume that the particle
is prepared with some initial momentum $p$. This means
in practice a very extended wavepacket with a very small
dispersion in momentum. We turn on at $t=0$ a step
of height~$V_{\tbox{step}}$. After a short time~$t$ we observe
that the classical phase space distribution is torn into
three pieces (see Fig.1):
phase space points that remain on the left side of the step;
phase space points that have crossed the step from left to right;
and phase space points that were all the time in
the right side of the step. The jump in the kinetic energy
of those points that have crossed the step is
\be{0}
\delta E_{\tbox{cl}} = -V_{\tbox{step}}
\ee
Classically we have in phase space points that move
with the original kinetic energy, and another
set of points that have gone through an abrupt
change of kinetic energy. Thus the energy distribution
consists of two delta peaks. We would like to know
what is the corresponding energy distribution
in the quantum mechanical case.


A similar phase space picture emerges in the analysis
of the "moving wall" problem. As illustrated 
in Fig.~2 we have a particle of mass~$\mass$ 
and energy~$E$ bouncing back and forth inside 
a one dimensional box. One wall of the box 
is displaced with a velocity~$V_{\tbox{wall}}$,  
which is assumed to be much smaller compared with 
the velocity $v_{\tbox{E}}=(2E/\mass)^{1/2}$ 
of the bouncing particle. 
Consider an initial microcanonical distribution. 
After a short time~$t$ some of the phase space points
collide with the wall which is moving
with velocity $V_{\tbox{wall}}$. 
Consequently their velocity undergoes
a change~${v \mapsto -v+2V_{\tbox{wall}}}$,
and accordingly the energy jump is:
\be{0}
\delta E_{\tbox{cl}} = -2 \mass v_{\tbox{E}} V_{\tbox{wall}}
\ee
Thus after a short time the energy distribution
consists of two delta peaks: one corresponds to those
phase space points that did not collide with the moving wall,
and the other corresponds to those phase space points
that did collide with the moving wall.
We ask what is the corresponding quantum result.
Namely, how the probability is distributed
among the energy levels in the quantum mechanical case.
It is implicit that we are going to work in
the adiabatic (wall location dependent) basis,
else the question is mathematically ill defined.


Possibly the most interesting and experimentally relevant model
is that of a one-dimensional EMF-driven ring (Fig.3).
The classical analysis for this problem is very simple:
each time that the particle crosses the EMF step its
energy changes by
\be{0}
\delta E_{\tbox{cl}} =  e V_{\tbox{EMF}}
\ee
So also here we have energy jumps.
Surprisingly this problem is interesting
even if we do not add a scatterer.

\section{Beyond the Fermi golden rule, the semiclassical regime}

Both in the case of the ``moving wall" and in
the case of the driven ring we have after
a short time a finite probability to find the system
with a different energy. So we may say that there is
some finite probability to make a transition
\be{0}
E \ \ \longmapsto \ \ E + \delta E_{\tbox{cl}}
\ee
Going to the quantum mechanical problem we may
wonder whether or how we get from the Schrodinger
equation such transitions.  We are used to the
Fermi golden rule picture of transitions
\be{999}
E \ \ \longmapsto \ \ E + ``\hbar\omega"
\ee
where $\omega$ is the frequency of the driving.
But here we do not have periodic (``AC") driving
but rather linear (``DC") driving.
Moreover, $\delta E_{\tbox{cl}}$ is an $\hbar$-
independent quantity.
It turns out that indeed there exists
a regime where the dynamics 
is classical-like (Fig.4).   
This semiclassical regime 
is defined by the obvious condition
\be{1000}
\delta E_{\tbox{cl}} \ \ \gg \ \ \Delta
\ee
where $\Delta$ is the level spacing.
In the case of the ``moving wall" problem
this condition can be written as
\be{1001}
V_{\tbox{wall}} \gg \frac{\hbar}{\mass L}
\ee
where $L$ is the size of the box.
It is easily verified that this
condition is just the opposite
of the adiabatic condition.
The case of the EMF-driven ring is
somewhat richer. The condition that
defines the semiclassical regime becomes
\be{0}
V_{\tbox{EMF}} \gg \frac{\hbar v_{\tbox{E}}}{L}
\ee
where $L$ is the length of the ring.
It is easily verified that this
condition is just the opposite
of the diabaticity condition.
The diabatic regime is defined as that
where transitions between energy levels
of a ``free" ring can be neglected.
If there is a small scatterer inside
the ring a stronger condition than diabaticity
is required in order to maintain adiabaticity:
\be{0}
V_{\tbox{EMF}} \ll  (1-g)\frac{\hbar v_{\tbox{E}}}{L}
\ee
where $g\sim 1$ is the transmission of the
scatterer. The adiabatic regime is defined
as that where transitions between the actual
energy levels of the ring can be neglected.
This is the regime where the Landau-Zener mechanism
of transitions at avoided crossings \cite{locGT,wilk} becomes
significant. The three regimes in the EMF-driven
ring problem are illustrated in the diagram of Fig.5.

Our main interest is in the non-trivial
semiclassical regime as defined by Eq.(\ref{e1000}).
In order to reconcile our semiclassical intuition
with the quantum Fermi Golden rule picture
we have to assume that the quantum dynamics self-generates
a frequency ${"\hbar \omega " = \delta E_{\tbox{cl}}}$.
Indeed it has been argued in Ref.\cite{wld}
that the non-perturbative mixing of levels
on the small energy scales generate this
frequency, while the re-normalized transitions
on the large (coarse grained) energy scales
are FGR-like.
However, an actual mathematical analysis of the
dynamics has not been introduced, and was left
as an open problem.

\section{Particle pulsed by a step}

The simplest example for a semiclassical energy jump
is provided by the "step problem". The time
dependent Hamiltonian is:
\be{15}
{\cal H} =
\frac{p^2}{2\mass} +
\left\lbrace
\begin{array}{ll}
0 & t < 0 \\
V_{\tbox{step}} & t \geq 0
\end{array}
\right.
\ee
For this Hamiltonian ``energy space"
is in fact ``momentum space", so it is
more natural to refer to ``momentum jumps".
Obviously we can translate any small
change in momentum to energy
units via $\delta E = v_{\tbox{E}} \delta p$,
where $v_{\tbox{E}}=(2E/\mass)^{1/2}$
is the velocity of the particle
in the energy range of interest.

The phase space dynamics after kicking
an initial momentum state $p_0$ at $t=0$
is illustrated in  Fig.1b.
It is clear that the emerging momentum distribution is
\be{16}
\rho_t(p) \ \ = \ \ \left[1-\frac{v_{\tbox{E}} t}{L}\right]\delta\Big(p-p_0\Big)
+  \frac{v_{\tbox{E}} t}{L} \delta\Big(p-(p_0+\delta p_{\tbox{cl}})\Big)
\ee
where $\delta p_{\tbox{cl}} \ \ = \ \  -V_{\tbox{step}}/v_{\tbox{E}}$,
and $L$ is the spatial extent of the wavepacket.
From here on we set $L=1$ as implied by the standard
density normalization of the momentum state $\eexp{ip_0x}$.
It is implicit in the following analysis that we assume
a very extended wavepacket ($v_{\tbox{E}}t \ll L$).
The emerging momentum distribution can be characterized by
its moments with respect to $p=p_0$. Namely:
\be{17}
\langle (p-p_0)^r \rangle
\ \ = \ \ \delta p_{\tbox{cl}}^r \times v_{\tbox{E}} t
\ \ \ \ \ \ \ \ \mbox{$r=1,2,3,4,...$}
\ee
All the moments are finite and grow linearly with time.
Below we are going to derive the quantum result.
Omitting the trivial ${\delta(p-p_0)}$ term,
and the back reflection term, the final result
for the forward scattering is
\be{18}
\rho_t(p) \ \ = \ \
|\bra p | {\cal U} | p_0 \ket|^2
\ \ = \ \ \frac{\delta p_{\tbox{cl}}^2 }{(p-p_0)^2} v_{\tbox{E}}^2 t^2
\ \sinc^2\left[\frac{1}{2}
\left(p-(p_0+\delta p_{\tbox{cl}})\right) v_{\tbox{E}} t \right]
\ee
for which
\be{0}
\langle (p-p_0)^r \rangle \ \ = \ \
\left\{
\begin{array}{ll}
\delta p_{\tbox{cl}} \times v_{\tbox{E}} t 
- \sin(\delta p_{\tbox{cl}} v_{\tbox{E}} t)  
\ \ \ \ \ \ \
& \mbox{for $r=1$} \\
\delta p_{\tbox{cl}}^2 \times v_{\tbox{E}} t
& \mbox{for $r=2$} \\
\infty
& \mbox{for $r>2$}
\end{array}
\right.
\ee
Let us compare the energy distribution in the classical and
quantum-mechanical cases (Fig.~7). As the
time~$t$ becomes much larger than $\hbar/V_{\tbox{step}}$ the
semiclassical peak is resolved. But we never get detailed QCC,
because all the high (${r>2}$) moments of the distribution diverge.

It should be appreciated that the power law tails that we get 
here for the energy distribution are the ``worst case" that 
can be expected. They emerge because the phase space distribution 
is torn in the momentum direction. In space representation 
this reflects a discontinuity in the derivative of the wavefunction. 
This explains why the tails go like $1/p^4$.
We are going to encounter the same type 
of power law tails also in the other examples.

\subsection{Derivation of the quantum result:}

The rest of this section is devoted for the derivation of the
quantum result and can be skipped in first reading. The momentum
states are denoted as $|p\rangle$. In order to simplify the
calculation we approximate the dispersion relation, within the
energy window of interest, as linear $E = v_{\tbox{E}} k$. This
implies that back-reflection is neglected. Once the step is turned
on, $|p\rangle$ are no longer the stationary states. The new
stationary states are
\be{0}
|k \rangle \ \ \longmapsto \ \
\Theta(-x) \eexp{ikx} + \Theta(x)\eexp{i(k+u)x}
\ee
where we use the notation ${u=\delta p_{\tbox{cl}}}$.
Note that these form a complete orthonormal set in the
sense ${\bra k_1 | k_2 \ket = 2 \pi\delta(k_1-k_2)}$.
The transformation matrix from the old to the new basis is
\be{55}
\bra p | k \ket
&=& \int^0_{-\infty} \eexp{-i  (p-k) x}dx
+ \int^{\infty}_0 \eexp{-i  (p-k-u) x}dx
\\
&=&
\pi \delta(p-k)+\frac{i } {p-k}
+\pi\delta(p-k-u)-\frac{i } {p-k-u}
\ee
Before we go on with the calculation we note 
that the following elementary integral 
can be found in any mathematical handbook: 
\be{-1}
\int_{-\infty}^{\infty}
\frac{dk}{2\pi}
\frac{1}{(k - p_2)(k - p_1)} 
\eexp{ikt}
\ \ = \ \ 
\frac{i }{2(p_2-p_1)} \left(\eexp{i  p_2 t}
- \eexp{i  p_1 t}\right)
\ \ \ \ \ \ \mbox{$p_2 {\ne} p_1$, $t>0$}
\ee
We notice that the result on the RHS if finite for $p_2=p_1$ 
while in fact it should diverge. This suggests that 
there is a missing delta term ${C \eexp{i  p_2 t}\delta(p_2-p_1)}$
where $C$ is a constant. In order to find this constant 
we have regularized this Fourier integral: 
\be{-1}
\lim_{\delta\rightarrow0}
\int_{-\infty}^{\infty}
\frac{dk}{2\pi}
{\frac{k-p_2}{(k-p_2)^2+\delta^2}\frac{k-p_1}{(k-p_1)^2+\delta^2}}
\eexp{ikt} \hspace*{3cm}
\\ \nonumber
\ \ = \ \ 
\frac{i }{2(p_2-p_1)} \left(\eexp{i  p_2 t}
- \eexp{i  p_1 t}\right)
+ \frac{\pi}{2} \eexp{i  p_2 t}\delta(p_2-p_1)
\ \ \ \ \ \ \mbox{$t>0$}
\ee
With the above we can calculate the matrix elements
of the evolution operator:
\be{-1}
\bra p_{} | {\cal U} | p_0\ket
&=& \sum_k {\eexp{-i  E_kt}\bra p_{}|k\ket \bra k|p_0\ket}
= \frac{1}{2\pi} \int_{-\infty}^{\infty}
\eexp{-i   v_{\tbox{E}} kt}\bra p_{}|k\ket \bra k|p_0\ket dk
\\ \nonumber
&=& \pi \delta(p_{}-p_0)\left(\eexp{i 
 v_{\tbox{E}} p_{} t}+\eexp{i   v_{\tbox{E}} (p_{}+u)t}\right)
\\ \label{57}
&+& \frac{i 
u}{(p_{}-p_0)(p_{}-p_0-u)}\left(\eexp{i   v_{\tbox{E}} (p_0+u)t} -
\eexp{i   v_{\tbox{E}} p_{} t}\right)
\ee
We have ${\bra p_{} | {\cal U}(t=0) | p_0\ket = 2\pi \delta(p_{}-p_0)}$
as required. The interesting part of this expression is the second
terms which is non vanishing for $p\ne p_0$. Taking its absolute value
and squaring we get after some algebra  Eq.(\ref{e18}).

\section{Particle in a box with a moving wall}

\subsection{The Schrodinger equation:}

We consider a particle in an infinite well.
The left wall is assumed to be fixed at $x=x_0$,
while the right wall at $x=X(t)$ is moving
with constant velocity $\dot{X}=V_{\tbox{wall}}$.
The size of the box is $L(t)=X(t)-x_0$.
Classically the dynamics is very simple:
each time that the particle hits the moving wall
its energy jumps by
$\delta E_{\tbox{cl}}=2\mass v V_{\tbox{wall}}$.
In the quantum mechanical case we work in the adiabatic
basis. The adiabatic energy levels and the eigenstates
for a given value of $L$ are:
\be{0}
E_n = \frac{1}{2\mass} \left(\frac{\pi \hbar}{L} n\right)^2
\ee
\be{0}
\Psi^{(n)}(x) = (-1)^n \sqrt{\frac{2}{L}}
\sin{\left(\frac{\pi n}{L}x\right)}
\ee
We use the standard prescription in order
to write the Schrodinger equation in the adiabatic basis.
Using the notations of Ref.\cite{pmc}
the equation is written as
\be{79}
\frac{d a_n}{d t} = -\frac{i }{\hbar}E_n a_n  +
i  \dot{X} \sum_m{A_{nm} a_m}
\ee
where
\be{80}
A_{nm} = i  \bra \Psi^{(n)} \Big| \pd{}{X} \Psi^{(m)} \ket
\ee
Hence, the Schrodinger equation for the problem in the adiabatic
basis is \cite{doescher,prm}:
\be{20}
\frac{da_n}{dt} = - \frac{i}{\hbar} E_n a_n - \frac{V_{\tbox{wall}}}{L}
\sum_{m (\ne n)} \frac{2nm}{n^2-m^2} \ a_m
\ee
%

\subsection{The generated dynamics:}

Let us assume that the initial preparation
is $a_n(0)=\delta_{nm}$.
The mean level spacing for the 1D box is
$\Delta = \pi\hbar v_{\tbox{E}}/L$.
If $\delta E_{\tbox{cl}} \ll \Delta$ one finds out,
by inspection of Eq.(\ref{e20}),
that the dynamics is adiabatic,
meaning that $a_n(t)\sim\delta_{nm}$.
On the other hand, if $\delta E_{\tbox{cl}} \gg \Delta$,
one expects to find a semiclassical transition
$E \mapsto E + \delta E_{\tbox{cl}}$.

How can we explain the $E \mapsto E + \delta E_{\tbox{cl}}$
transition from quantum-mechanical point of view?
For this purpose we can adopt the
core-tail picture of Ref.~\cite{frc}:
The `core' consists of the levels that are
mixed non-perturbatively; The `tail' is formed by
first order transitions from the core.
Originally this picture has been applied 
to analyze the energy spreading in ``quantum chaos" 
driven systems. Here, the (non-chaotic) moving 
wall problem allows a much simpler 
application~\cite{wld}. The analysis is carried 
out in two steps which are summarized below.

The first step in the ``core-tail" picture 
is to analyze the {\em parametric evolution}
which is associated with Eq.(\ref{e20}).
This means to solve Eq.(\ref{e20}) without
the first term in the RHS. (This is the so-called sudden limit).
Obviously the resultant $\tilde{a}_n(t)$
is a function of $\delta X = V_{\tbox{wall}}t$, while
$V_{\tbox{wall}}$ by itself makes no difference. The solution
depends only  on the endpoints $x(0)$ and $x(t)$. 
By careful inspection of Eq.(\ref{e20}) one observes
that a level is mixed with the nearby level whenever
the wall is displaced a distance $\lambda_{\tbox{E}}/2$, 
where $\lambda_{\tbox{E}}=2\pi\hbar/(\mass v_{\tbox{E}})$ is the 
de Broglie wavelength. The time scale which is associated
with this effect is obviously
\be{0}
\tau_{\tbox{qm}} = \frac{\lambda_{\tbox{E}}/2}{V_{\tbox{wall}}}
\ee
The second step is to analyze
the actual time evolution. This means to take into account
the effect of the first term in the RHS of Eq.(\ref{e20}),
and to understand how the resultant $a_n(t)$ differs
from $\tilde{a}_n(t)$. One observes that the `parametric' mixing 
of nearby levels modulates the transition amplitude.
The modulation frequency is
\be{220}
``\omega" =  \frac{2\pi}{\tau_{\tbox{qm}}}
\ee
Once combined with the FGR Eq.(\ref{e220}) it leads
to the anticipated semiclassical result Eq.(\ref{e999}).
It is not difficult to argue that 
the period of this semiclassical transition is  
\be{0}
\tau_{\tbox{cl}} = \frac{2L}{v_{\tbox{E}}} 
\ee
which is the time to make one round between 
the walls of the well. Since we are dealing 
with a simple 1D system this coincides with 
the Heisenberg time:
\be{0}
t_{\tbox{H}} = \frac{2\pi\hbar}{\Delta} = \tau_{\tbox{cl}} 
\ee
The ratio $\tau_{\tbox{cl}}/\tau_{\tbox{qm}}$ determines the number 
of nearby level transitions per period. 
Obviously the semiclassical condition Eq.(\ref{e1001}) 
requires this ratio to be much larger than unity. 
The disadvantage of the above heuristic picture is that 
it does not lead to a satisfactory quantitative results. 
Therefore, in later sections we discuss an optional route 
of analysis via a reduction to a tight binding model.

\subsection{Numerical Simulation}

The solution of Eq.(\ref{e20}) 
becomes very simple if we 
make the approximation ${L(t) \approx L_0}$. 
This holds as long as the wall displacement 
is small. We have verified that the 
associated numerical error is very small.
Using units such that ${L_0=\mass=\hbar=1}$ 
we define a diagonal matrix ${\bm{E} =\mbox{diag}\{ \pi^2n^2/2 \}}$   
and a non-diagonal matrix, 
${\bm{W}=\{-i 2\alpha nm/(n^2{-}m^2)\}}$ 
with zeros along the diagonal, 
and where ${\alpha=V_{\tbox{wall}}/L}$.
The evolution matrix in the adiabatic basis 
is obtained by exponentiation:  
\be{0}
\bm{U}(t) = \exp{\left[-i  \ t \ \left(\bm{E}+\bm{W}\right)\right]}
\ee
Fig.8 illustrates the time dependence of probability 
distribution $|a_n(t)|^2$ for a particle initially 
prepared at $n_0=50$.  Fig.8a displays the solution 
in the adiabatic regime: the particle stays at the 
same level. Fig.8b displays the solution
in the semiclassical regime: at each moment the particle 
partially stays at the same energy, and partially 
makes classical-like transition to the next energy strip.  
Fig.9 highlights the energy splitting of the wavepacket  
during the transition.

If we want to avoid the $L(t) \approx L_0$ approximation, 
the price is a time dependent $\bm{E}$ and  $\bm{W}$ matrices.
Then the calculation should be done in small $dt$ time steps:
\be{102}
\bm{U}(t) 
= 
\prod_{t'=dt}^{t}
\exp{\left[-i  \ dt \ \bm{W}(t') \right]}
\ \exp{\left[-i  \ dt \ \bm{E}(t') \right]}
\ee
The state of the system is described by 
a truncated column vector ${\bm{a}=\{a_n\}}$ 
of length~$N$. 
Optionally it is possible to represent 
the state of the system in the Fourier 
transformed basis. The elements $A_k$ 
of the Fourier transformed vector are 
labeled by $k=(2\pi/N)\tilde{n}$, 
where $\tilde{n} \ \mbox{mod}(N)$ is an integer.
The practical implementation of Eq.(\ref{e102})  
is greatly simplified if $\bm{W}_{nm}$ 
is a function of the difference ${n-m}$. 
In such case $\bm{W}$ is transformed into 
a diagonal matrix~$\tilde{\bm{W}}$.  
Consequently one can use the standard 
fast Fourier transform (FFT) algorithm in order 
to propagate a given state vector. Namely, 
\be{101}
\bm{a}(t) 
= 
\prod_{t'=dt}^{t}
\mbox{FFT}^{-1}
\ \exp{\big[-i  \ dt \ \tilde{\bm{W}}(t') \big]}
\ \mbox{FFT}
\ \exp{\big[-i  \ dt \ \bm{E}(t') \big]}
\ \bm{a}(0) 
\ee
where both $\bm{E}$ and $\tilde{\bm{W}}$ are diagonal.
In the moving wall problem $\bm{W}_{nm}$ is  
mainly proportional to $1/(n{-}m)$,
so the FFT method is applicable 
if we restrict the energy range of interest. 
In the next section we shall consider the EMF-driven ring 
problem, leading to a very similar evolution equation,  
where the FFT method is strictly applicable.

\section{Particle in an EMF-driven ring}

\subsection{The Schrodinger equation:}

We consider a 1D-ring driven by an EMF (Fig.3).
The EMF is induced by a time-dependent flux which is described
by the vector potential
\be{0}
A(x,t) = \Phi(t) \delta(x - x_0)
\ee
This means that the electric field is
\be{0}
\mathcal{E}(x) = V_{\tbox{EMF}} \delta(x-x_0)
\ee
where $V_{\tbox{EMF}}=-\dot{\Phi}=\const$.
The Hamiltonian that generates the dynamics is
\be{22}
\mathcal{H}(\Phi(t))
= \frac{1}{2\mass} \left(\hat{p} - \frac{e}{c} A(\hat{x},t)\right)^2
\ee
with periodic boundary conditions over $x$. 
The length of the ring is~$L$.

Classically the dynamics is very simple: each time that the particle
crosses $x=x_0$ its energy jumps by $\delta E_{\tbox{cl}} = eV_{\tbox{EMF}}$.
In the quantum mechanical case it is convenient to work in
the so-called diabatic basis. The diabatic energy levels for
a given value of $\Phi$ are
\be{0}
E_n = \frac{1}{2\mass} \left(\frac{2 \pi \hbar}{L} \right)^2
\left(n - \frac{\Phi}{\Phi_0} \right)^2
\ee
where $\Phi_0=2\pi\hbar c/e$. See Fig.4.
The Schrodinger equation that describes
the time evolution in the diabatic basis is
found using the same procedure as in the case
of the moving wall. We have to find the $A_{nm}$
as defined in Eq.(\ref{e80}) where now $X=\Phi$.
The only extra difficulty
is in finding the eigenstates $\Psi^{(n)}$
of  Eq.(\ref{e22}) because $A(x)$ depends
on $x$.  The calculation becomes much simpler
if we realize that they are related by a gauge
transformation to the eigenstates $\tilde{\Psi}^{(n)}$
of a much simpler Hamiltonian:
\be{73}
\tilde{\mathcal{H}}
= \frac{1}{2\mass} \left(p - \frac{e\Phi}{cL} \right)^2
\ee
Namely,
\be{70}
\Psi^{(n)}(x)
&=&
\exp{\left(\frac{i  e}{\hbar c} \Lambda(x)\right)}
\tilde{\Psi}^{(n)}(x)
\\
&=&
\exp{\left(\frac{i  e}{\hbar c} \Lambda(x)\right)}
\times
\frac{1}{\sqrt{L}}
\exp{\left(i  \frac{2 \pi n}{L}x\right)}
\\
&=&
\frac{1}{\sqrt{L}}
\exp{\left(i  \frac{2 \pi} {L} \left(\frac{\Phi}{\Phi_0}+n \right)x \right)}
\ee
where in the last line we set $x_0=0$ and the gauge function is 
\be{72}
\Lambda(x) = \frac{\Phi}{L} x
\ee
Using the above result we get
\be{0}
A_{nm} = -\frac{i }{\Phi_0}\frac{1}{n-m}
\ee
and accordingly
\be{23}
\frac{d a_n}{d t} = -\frac{i }{\hbar}E_n a_n
+\frac{V_{\tbox{EMF}}}{\Phi_0} \sum_{m(\neq n)}{ \frac{1}{n-m} a_m}
\ee

\subsection{The generated dynamics:}

The dynamics of an EMF-driven ring is very similar 
to the dynamics in the moving wall problem.
This is obvious from the phase space picture, and 
also by inspection of the equation for $a_n(t)$. 
Also the ``core tail" heuristic picture of section~5.2 
is easily adapted. The parametric scale that signifies 
mixing of nearby levels is now $\delta X = \Phi_0$ 
instead  $\delta X = \lambda_{\tbox{E}}/2$ leading to 
the quantum time scale 
\be{0}
\tau_{\tbox{qm}} = \frac{\Phi_0}{V_{\tbox{EMF}}}
\ee
The classical period is 
\be{0}
\tau_{\tbox{cl}} = \frac{L}{v_{\tbox{E}}}
\ee
and the semiclassical condition can be written 
as $\tau_{\tbox{qm}} \ll \tau_{\tbox{cl}}$.

\hide{
From the experimental point of view it is possible to measure this kind of
transitions. For example, for a mesoscopic golden ring of the diameter of 3
microns we estimate the Heisenberg time as $t_{\tbox{H}} =
2\pi\hbar/\Delta = L/v_{\tbox{F}}$, where the Fermi velocity for
the gold is known from the literature $v_{\tbox{F}} = 1.6 * 10^6$ m/s. The mean
level spacing $\Delta$, therefore, is about $6.1 * 10^{-4}$ eV and it can be
resolved in an experiment.}

Since the energies are time dependent we have 
to use Eq.(\ref{e102}) for the calculation of the
time evolutions. Furthermore, $\bm{W}$ 
is diagonal in the momentum representation, 
and therefore we can use the FFT method Eq.(\ref{e101}) 
with $\tilde{\bm{W}}=\mbox{diag}\{-\alpha \ (k-\pi)\}$, 
where $k=(2\pi/N)\tilde{n}$ is defined $\mbox{mod}(2\pi)$. 
The results of the simulations are presented in Fig.10a 
and Fig.11. We shall further discuss these results 
in the next sections.

\section{Bloch electrons in a constant electric field (I)}

If we focus our interest in small energy interval, then
in both cases (moving wall, driven ring) the Schrodinger
equation in the adiabatic basis is approximately the same
as that of an electron in a tight binding model,
where $n$ is re-interpreted as the site index:
\be{24}
\frac{d a_n}{d t} = -i  E_n a_n
+ \alpha \sum_{m(\neq n)}{ \frac{1}{n-m} a_m}
\ee
with $E_n=\varepsilon n$. We use from here on $\hbar=1$ units.  
The scaled rate of the driving $\alpha$ is re-interpreted as 
the hopping amplitude between sites, 
while the levels spacing $\varepsilon$
is re-interpreted as an electric field.  
Assuming that the electron is initially at the site $n_0$ we would
like to find out what is the probability distribution
\be{0}
\rho_t(n) = |a_n(t)|^2
\ee
It is obvious that the adiabatic regime $\alpha \ll \varepsilon$ corresponds
to a large electric field that localizes the electron at
its original site. In the other extreme ($\alpha \gg \varepsilon$), 
if the effect of~$\varepsilon$ could have been ignored, 
we would observe unbounded Bloch ballistic motion.  
The effect of finite~$\varepsilon$ is to turn this 
motion into Bloch oscillations. We shall find below that 
the electron performs periodic motion which
we illustrate in Fig.~6: While the wavepaket drifts
with the electric field to the right, it shrinks
and disappears, and at the same time re-emerges on the left.
If we run  the simulation as a movie, it looks as if the
motion is from left to right. Still it is bounded in space
due to this ``re-injection" mechanism.

First of all we solve the equation for $\varepsilon=0$.
The Hamiltonian is diagonal in the momentum basis $k$
and therefore the general solution is
\be{0}
a_n(t)
= \sum_k A_k \eexp{i(kn-\omega_kt)}
= \int_{0}^{2\pi}\frac{dk}{2\pi} A_k \eexp{i(kn-\omega_kt)}
\ee
The dispersion relation is found by
transforming the Hamiltonian to the $k$ basis:
\be{82}
\omega_k
\ \ = \ \ i\alpha \sum_{n (\ne m)} \frac{\eexp{-i  k (n-m)}}{n-m}
\ \ = \ \ \alpha \ [\pi-k]
\ee
If we place at $t=0$ an electron at site $n_0$,
then $a_n=\delta_{n,n_0}$, and hence $A_k=\eexp{-ik n_0}$.
Then we get
\be{28}
a_n(t) = \frac{\sin{\pi \alpha t}}{\pi (\alpha t + n - n_0)}
\ee
Turning to the general case with $\varepsilon\ne0$
we substitute $a_n(t)=c_n(t)\eexp{-i  E_nt}$ and get
the equation
\be{88}
\frac{dc_n}{d t} =
\alpha \sum_{m(\neq n)}
\frac{\eexp{i  (n-m) \varepsilon t}}{n-m} c_m
\ee
This more complicated equation is still diagonal in the $k$ basis:
\be{0}
\frac{dC_k}{d t} =  -i\omega_k(t)  C_k
\ee
where
\be{0}
\omega_k(t) \ \  = \ \ \alpha \ [\pi-\mbox{mod}(k+\varepsilon t,2\pi)]
\ee
and its solutions is
\be{-1}
C_k(t)  &=& 
C_k(t{=}0)
\ \exp{\left[-i\int_0^t \omega_k(t') dt'\right] } 
\\ \nonumber
&=&   
\left\{
\begin{array}{ll}
\eexp{-i  k n_0 + i  \alpha
\left((k - \pi) t + \frac{\varepsilon t^2}{2}  \right)} 
&
\mbox{for $\varepsilon>0$ and $0<k<\overline{k}$}
\\
\eexp{-i  k n_0 + i  \alpha
\left((k - 3\pi)t + \frac{\varepsilon t^2}{2}  +
\frac{2\pi(2\pi-k)}{\varepsilon}\right)}
&
\mbox{for $\varepsilon>0$ and $\overline{k}<k<2\pi$}
\\
\eexp{-i  k n_0 + i  \alpha
\left((k - \pi) t + \frac{\varepsilon t^2}{2} +
\frac{2\pi k}{\epsilon} \right)}
&
\mbox{for $\varepsilon<0$ and $0<k<\overline{k}$}
\\
\eexp{-i  k n_0 + i  \alpha
\left((k + \pi)t + \frac{\varepsilon t^2}{2}
\right)}
&
\mbox{for $\varepsilon<0$ and $\overline{k}<k<2\pi$}
\end{array}
\right.
\ee
which is valid for $0<t<2\pi/|\varepsilon|$ 
and should be continued periodically in time. 
We have used the notation ${\overline{k} = -\varepsilon t \ \mbox{mod}(2\pi)}$. 
Now we can go back to position representation:
\be{98}
c_n(t) 
=\int_{0}^{\overline{k}(t)} \frac{dk}{2\pi} C_k \eexp{ikn}
+\int_{\overline{k}(t)}^{2\pi} \frac{dk}{2\pi} C_k \eexp{ikn}
\ee
Taking the absolute value and squaring we get 
the following result for the probability distributions:
\be{100}
\rho_t(n) = \left(2\frac{\alpha}{\varepsilon}\right)^2
\frac{\sin^2\left(\frac{1}{2} \varepsilon t
\left(n-n_0+\alpha(t-\frac{2\pi}{|\varepsilon|})\right)\right)}
{(n-n_0+\alpha t)^2(n-n_0+\alpha(t-\frac{2\pi}{|\varepsilon|}))^2}
\ee
The above formula is valid for $0<t<2\pi/|\varepsilon|$ 
and it should be continued periodically in time.  
Fig.6 and Fig.10a illustrate the dynamics 
both schematically and numerically.
In the next section we further 
discuss the nature of this dynamics.

\section{Bloch electrons in a constant electric field (II)}

In order to appreciate the significance of the $\propto 1/(n{-}m)$ 
hopping we solve again the problem of Bloch electrons  
in a constant electric field, but this time with the 
``conventional" nearest neighbor hopping:
\be{0}
\frac{d a_n}{d t} 
= -i  E_n a_n 
+ \frac{\alpha}{2} [a_{n+1} - a_{n-1}]
\ee
with $E_n=\varepsilon n$. The initial preparation 
at $t{=}0$ is $a_n = \delta_{n,n_0}$.
We substitute $a_n = \eexp{-i  E_n t}c_n$ and get the equation:
\be{0}
\frac{d c_n}{d t} =
\frac{\alpha}{2} (\eexp{-i  \varepsilon t}c_{n+1}-\eexp{i  \varepsilon t}c_{n-1})
\ee
This equation becomes diagonal in the $k$ basis:
\be{0}
\frac{dC_k}{d t} =  -i\omega_k(t)  C_k
\ee
where
\be{0}
\omega_k(t) \ \  = \ \ \alpha \sin(\varepsilon t + k)
\ee
Its solutions is
\be{0}
C_k(t) \ \ = \ \
C_k(t = 0)\times
\exp{\left[-i\int_0^t \omega_k(t') dt'\right] }
\ee
Solving the above integral and making the inverse Fourier transform we obtain:
\be{0}
c_n(t) = J_{n-n_0}\left(\frac{2\alpha}{\varepsilon}
\sin\left(\frac12 \varepsilon t\right) \right)
\ee
where $J()$ is the Bessel function of the first kind.
Taking the absolute value and squaring we get the probability distribution:
\be{0}
\rho_t(n) = \left|J_{n-n_0}\left(\frac{2\alpha}{\varepsilon}
\sin\left(\frac12\varepsilon t\right) \right)\right|^2
\ee
Fig.10b illustrates the dynamics. As in the previous 
problems we can distinguish between two time scales.
One is related to the diagonal part of the Hamiltonian, 
and the other one to the hopping term. Keeping the 
same notations as in previous sections these are 
\be{0}
\tau_{\tbox{cl}} &=& 2\pi/\varepsilon \\
\tau_{\tbox{qm}} &=& 1/\alpha 
\ee
The nature of the dynamics in the case of 
$\propto 1/(n{-}m)$ hopping and in the case 
of near neighbor hopping is quite different, 
as it can be appreciated by comparing Fig.10a and Fig.10b. 
This is related to the additional symmetries 
in the latter case. In order to explain this point 
let us use the notation $\bm{U}(\alpha,\varepsilon)$
that emphasizes that the evolution depends 
on two parameters, the first one is associated 
with the kinetic term $\bm{W}=w(\hat{p})$ 
and the other one with the potential term $\bm{E}=\epsilon(\hat{x})$. 
For clarity we use $\hat{x}$ for the position 
coordinate and $\hat{p}$ for the quasi-momentum. 
In both cases we have the anti-unitary symmetry 
${(x,p)\mapsto (x,-p)}$, that maps $\bm{E}$ to $\bm{E}$
and $\bm{W}$ to $-\bm{W}$. 
Consequently $\bm{U}(\alpha;\varepsilon)$ is mapped 
to $\bm{U}(\alpha;-\varepsilon)$. 
This implies that the spreading does not depend 
on the direction of the electric field. 
This is a peculiarity of tight binding models.
The conventional time reversal symmetry, 
for which the kinetic term $\bm{W}$ is left invariant,  
is ${(x,p)\mapsto (x,\pi{-}p)}$. This symmetry  
characterizes the near neighbor hopping, but not 
the $\propto 1/(n{-}m)$ hopping.   
This symmetry implies that the spreading 
looks the same if we reverse the signs of both 
$\alpha$ and $\varepsilon$, which is like 
reversing the time. If we combine the two symmetries 
we deduce that the dynamics, in the case of 
the near neighbor hopping, should be indifferent 
to the sign of $\alpha$. Note that the 
combined symmetry that leads to this conclusion 
is the unitary mapping  ${(x,p)\mapsto (x,p{+}\pi)}$.    
Thus, in both cases [$\propto 1/(n{-}m)$ hopping and near neighbor hopping]  
we have generalized Bloch oscillations, but in the former 
case they are unidirectional (Fig.6), while in the latter 
case they are bi-directional.

\section{Discussion}

Within Linear response theory (LRT) the energy absorption
of a quantum system is determined by the
correlation function of the perturbation term.
In general one can argue that there is a very
good QCC for the correlation functions,
and hence one expects {\em restricted} QCC
in the energy absorption process.
The persistence of {\em restricted} QCC
in the $t\rightarrow\infty$ limit requires
the additional assumption of having
a coarse grained Markovian-like behavior
for long times. Depending of the
context one should further assume that the
environment supplies both weak decoherence effect
that makes the break time $t^*$ irrelevant,
and a weak relaxation effect so as to achieve
a steady state. Then it is possible to use
the same argumentation as in the derivation
of the central limit theorem in order to argue
that all the higher moments become Gaussian-like.

Thus, the common perception is that the leading result
for the response of a driven system should be the same
classically and quantum mechanically. For example,
such is the case if one calculates the conductance of
a diffusive ring \cite{kamenev}: The leading order result
is just the Drude expressions, and on top there
are weak localization corrections.

The above reasoning illuminates that the long time
response is based on the short time analysis.
Moreover, one realizes that the second moment
of the evolving energy distribution has
a special significance. Still, all the above
observations are within a very restrictive framework
of assumptions. In practice it is of much interest
to explore the limitations of LRT, and to obtain
a more general theory for response.

The theory for the response of closed isolated driven
quantized chaotic mesoscopic systems is far from
being trivial, even if the interactions between the particles
are neglected. In the case of a generic quantized
chaotic systems two energy scales are involved:
the mean level spacing $\Delta \propto \hbar^d$,
where $d=2,3$ is the dimensionality of the system,
and the semiclassical energy scale $\hbar/\tau_{cl}$.
It is implied (see the mini-review of Ref.\cite{dsp}) that
there are generically three regimes depending on
the rate $\dot{X}$ of the driving:
\begin{itemize}
\item The adiabatic (Landau Zener) regime
\item The Fermi-golden-rule (FGR, Kubo) regime
\item The semiclassical (non-perturbative) regime
\end{itemize}
Most of the literature in mesoscopic physics
is dedicated to the study of the dynamics in
either the adiabatic or the FGR regimes.
The existence of a non-perturbative regime \cite{crs,dsp}
is not yet fully acknowledged, though it has been
established numerically in the RMT context \cite{rsp}.

Driven one-dimensional systems are non-generic
because typically the semiclassical energy scale
coincides with the mean level spacing. In other words:
the Heisenberg time $t_H=2\pi\hbar/\Delta$ is
the same as the classical time $\tau_{cl}$
rather than being much larger. Indeed we have seen that
in the ``moving wall" problem we have just two regimes:
the adiabatic regime and the semiclassical regime.

The EMF-driven ring is a prototype problem
in mesoscopic physics. It is richer than the
``moving wall" problem because a small scatterer
introduces a very small energy scale,
the level splitting, and hence we have three
regimes rather than two: adiabatic, diabatic and semiclassical.

The semiclassical regime in the study of EMF-driven rings
has not been explored so far. One important observation
is that contrary to LRT the gauge of the vector potential {\em does matter}.
Most of past studies assume that the vector potential
is $A(x,t)=\Phi(t)/L$. It is true that in LRT
the same result for the conductance
is obtained with  $\tilde{A}(x,t)=\Phi(t)\delta(x-x_0)$.
If we try to go from $\tilde{A}(x,t)$ to $A(x,t)$
using a guage transformation, the ``price" is
a modified $V(x)$ that features a linear ramp with
a step-like drop at $x=x_0$. This modification of $V(x)$
can be neglected only in the LRT regime.
The semiclassical condition of Eq.(\ref{e1000})
is just that opposite of this LRT requirement.

The semiclassical dynamics implies diffractive 
energy spreading. The mixing of levels in the small energy 
scales induces jumps in energy space.  
The realization that this diffractive energy spreading 
can be re-interpreted using a tight binding Bloch model 
follows in spirit the celebrated reduction \cite{qkr}
of dynamical localization in periodically kicked
systems to a tight binding Anderson problem. An interesting
feature is the hopping that goes like $\propto 1/(n-m)$.
This hopping leads to an unidirectional rather than 
bidirectional Bloch oscillations, 
as implied by the semiclassical reasoning.

\section{Appendix: The robustness of restricted QCC}

The simplest way to illuminate the robustness
of the second moment is by adopting
a heuristic phase space picture language.
Given two operators $\hat{A}=A(\hat{x},\hat{p})$ and $\hat{B}=B(\hat{x},\hat{p})$,
with Wigner Weyl representation $A_{\tbox{WW}}(x,p)$ and $B_{\tbox{WW}}(x,p)$
we have the exact identity
\be{0}
\trc(\hat{A}\hat{B}) = \int \frac{dxdp}{2\pi}
A_{\tbox{WW}}(x,p) B_{\tbox{WW}}(x,p)
\ee
If we can justify the replacement
of $A_{\tbox{WW}}(x,p)$ by $A(x,p)$
and $B_{\tbox{WW}}(x,p)$ by $B(x,p)$
then we get QCC.
The rule of the thumb is that in order to justify such
an approximation the phase space contours of $A(x,p)$
and $B(x,p)$ should be significantly different.
Otherwise the transverse structure of the Wigner-Weyl
functions should be taken into account.
This reasoning can be regarded as a phase space
version of the stationary phase approximation.

Let $\hat{A}=[H(\hat{x},\hat{p})]^r$ the $r$th
power of the Hamiltonian $\mathcal{H}=H(\hat{x},\hat{p})$,
and let $\hat{B}=\rho(H_0(\hat{x},\hat{p}))$
a stationary preparation
with the Hamiltonian $\mathcal{H}_0=H_0(\hat{x},\hat{p})$.
In such case $\trc(\hat{A}\hat{B})$
is the $r$th moment $\langle \mathcal{H}^r \rangle$
of the energy. If $H=H_0+\lambda V$,
and $\lambda$ is not large enough,
then we do not have detailed QCC.
This is discussed throughly in Ref.\cite{lds}.
But at the same time, irrespective of $\lambda$,
restricted QCC is robust.
The reason is that for the first two moments
we have the identities
\be{0}
\langle \mathcal{H} \rangle &=&
\langle \mathcal{H}_0 \rangle + \langle \hat{V} \rangle
\\
\langle \mathcal{H}^2 \rangle &=&
\langle \mathcal{H}_0^2 \rangle
+ 2\langle\mathcal{H}_0 \rangle \langle \hat{V} \rangle
+ \langle \hat{V}^2 \rangle
\ee
Thus the calculation of $\trc(\hat{A}\hat{B})$
with $\hat{A}=[H(\hat{x},\hat{p})]^r$ reduces
to the calculation of $\trc(\hat{A}\hat{B})$
with $\hat{A}=[V(\hat{x},\hat{p})]^r$.
We assume that $V$ and $H$ are not related
in any special way. It follows that
we have robust QCC for all the moments
of $V$, and consequently also for the
first two moments of $H$, irrespective of $\lambda$.

In order to generalize the above reasoning to
time dependent Hamiltonians, it is convenient
to adopt the Heisenberg picture.
Given that the system is prepared in a stationary state at $t=0$,
one can prove that
\be{0}
\langle\mathcal{H}(t)^2\rangle_0-\langle\mathcal{H}(0)^2\rangle_0
= \langle(\mathcal{H}(t)-\mathcal{H}(0))^2\rangle_0
\ee
where $\mathcal{H}(t)$ is the Hamiltonian $\mathcal{H}(X(t))$
in the {\em Heisenberg picture}. Such relation cannot not
be generalized to higher moments because of lack of
commutativity. Using
\be{0}
\frac{d\mathcal{H}}{dt}
\ \ = \ \frac{\partial\mathcal{H}}{\partial t}
\ \ = \ \ \dot{X}\hat{V}(t)
\ee
where $V \equiv \partial \mathcal{H} / \partial X$,
we can express ${ \langle(\mathcal{H}(t)-\mathcal{H}(0))^r\rangle_0 }$
as an integral over the correlation functions of the perturbation~$V(t)$.
The QCC for these correlation functions is robust,
and hence the QCC for the second moment is also robust.

\clearpage

\ack

We thank Tsachy Holovinger for preparing the first version
of the ``moving wall" simulation. AS thanks V. Goland for
fruitful discussions that helped solving the "Bloch electrons"
problem. The research was supported
by the Israel Science Foundation (grant No.11/02),
and by a grant from the DIP, the Deutsch-Israelische Projektkooperation.



\clearpage

\mpg{

\putgraph[width=0.5\hsize]{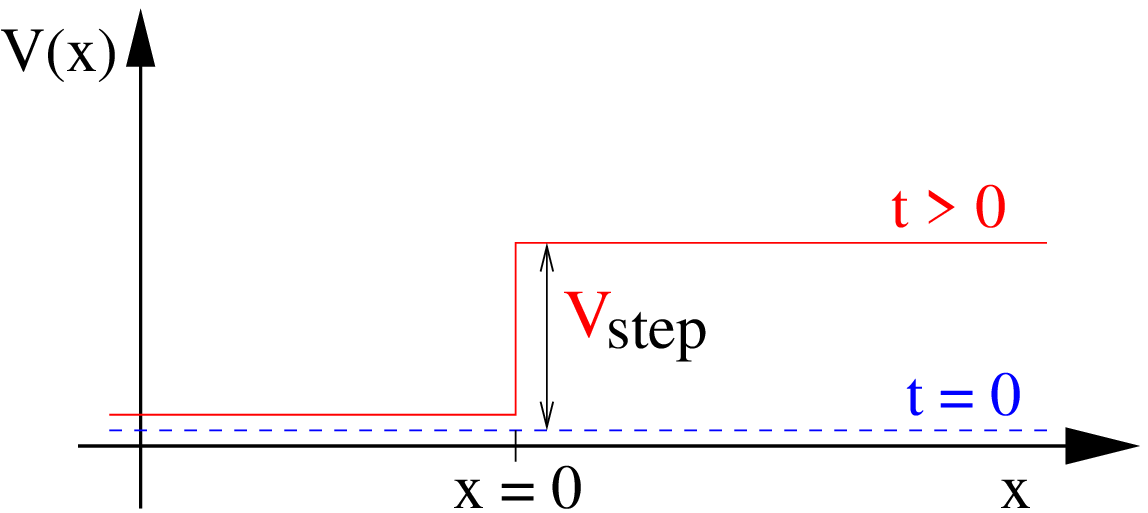}
\hfill
\putgraph[width=0.3\hsize]{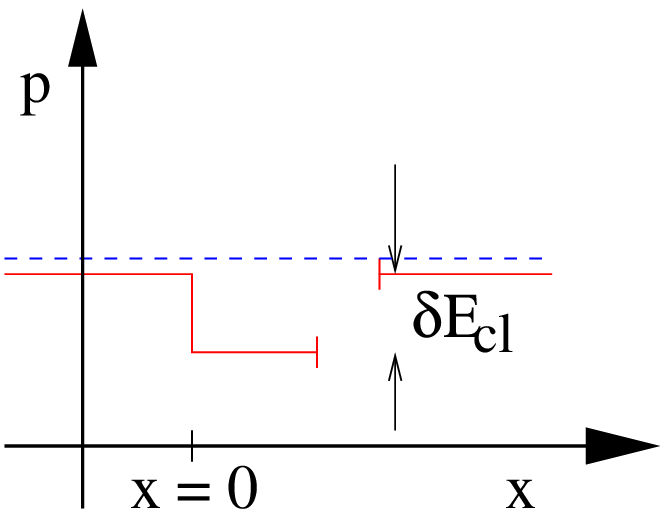}

{\footnotesize {\bf Fig.1:}
(a) Left panel: Picture of the potential. Before $t{=}0$ the potential is zero
(dashed line). After~$t{=}0$ the potential is the step function (solid line).
(b) Right panel: Phase space picture. Before $t{=}0$ there is no potential and the
momentum is constant (dashed line). The piece of the distribution that has 
passed $x{=}0$ after $t{=}0$ is boosted with $\delta E_{\tbox{cl}} = -V_{\tbox{step}}$.}

\ \\

\putgraph[width=0.5\hsize]{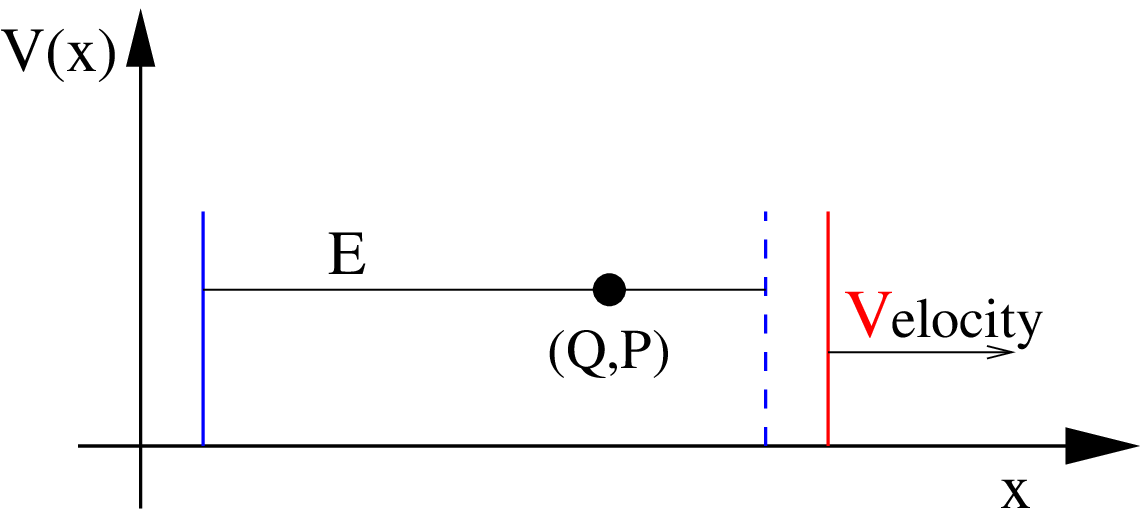}
\hfill
\putgraph[width=0.3\hsize]{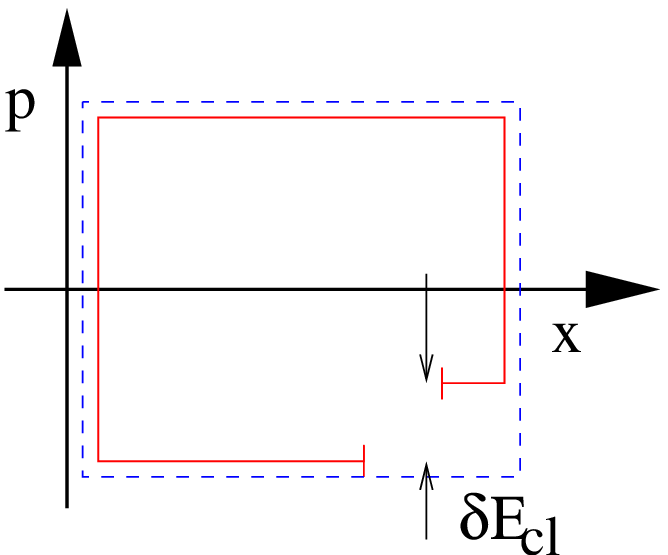}

{\footnotesize {\bf Fig.2:}
(a) Left panel: Potential well. The right wall is moving 
with a constant velocity $V_{\tbox{wall}}$.
(b)~Right panel: Phase space picture. If the wall were not moving, 
the distribution would evolve along the dashed line.  
If $V_{\tbox{wall}}$ is non zero, an energy 
jump $\delta E_{\tbox{cl}} = -2\mass v_{\tbox{E}} V_{\tbox{wall}}$
is associated with the collision, and one obtains 
the distribution which is illustrated by the solid line.}

\ \\

\putgraph[width=0.25\hsize]{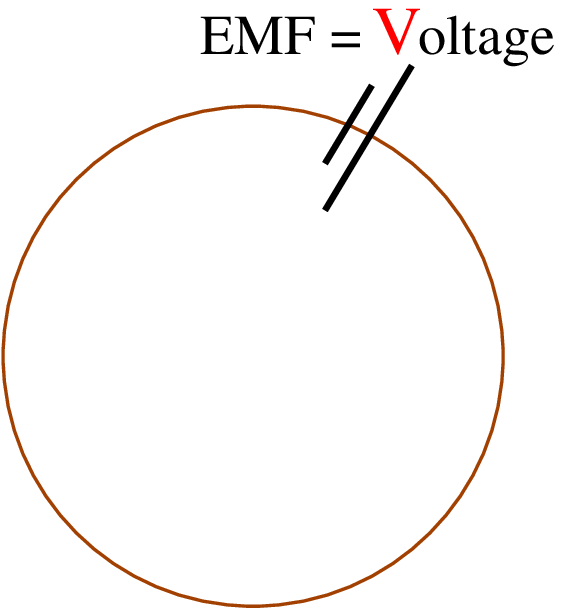}
\hfill
\putgraph[width=0.35\hsize]{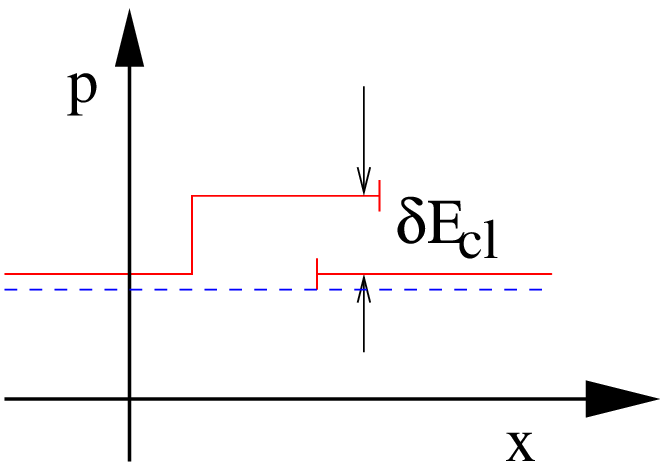}

{\footnotesize {\bf Fig.3:}
(a) Left panel: Ring with EMF.
(b) Right panel: Phase space picture. Without the EMF 
the momentum is a constant of the motion (dashed line). 
Else an energy jump $\delta E_{\tbox{cl}} = eV_{\tbox{EMF}}$
is associated with each crossing of the EMF step.  
The emerging phase space distribution is illustrated 
by the solid line.}

}

\mpg{

\putgraph[width=0.45\hsize]{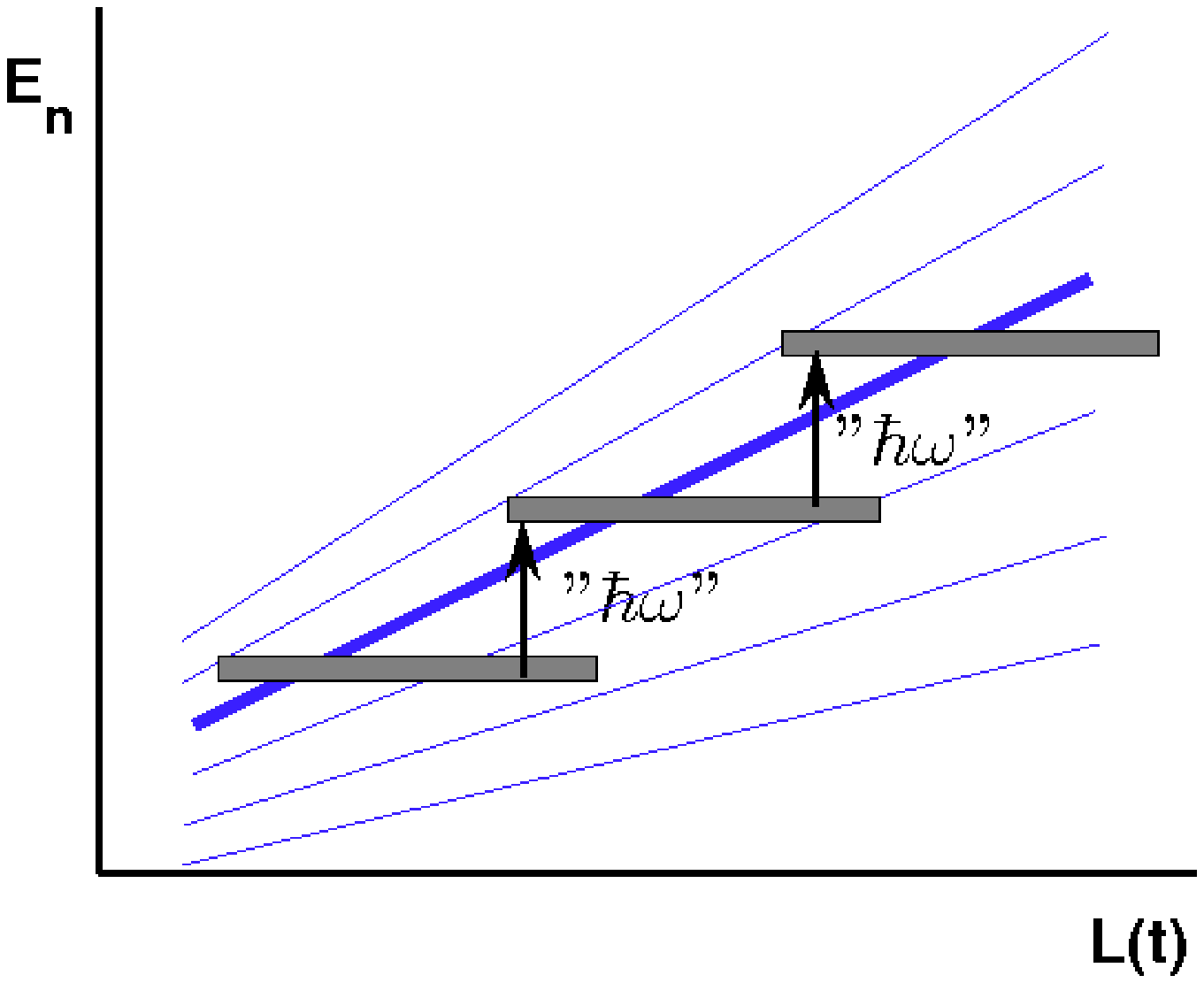}
\hfill
\putgraph[width=0.45\hsize]{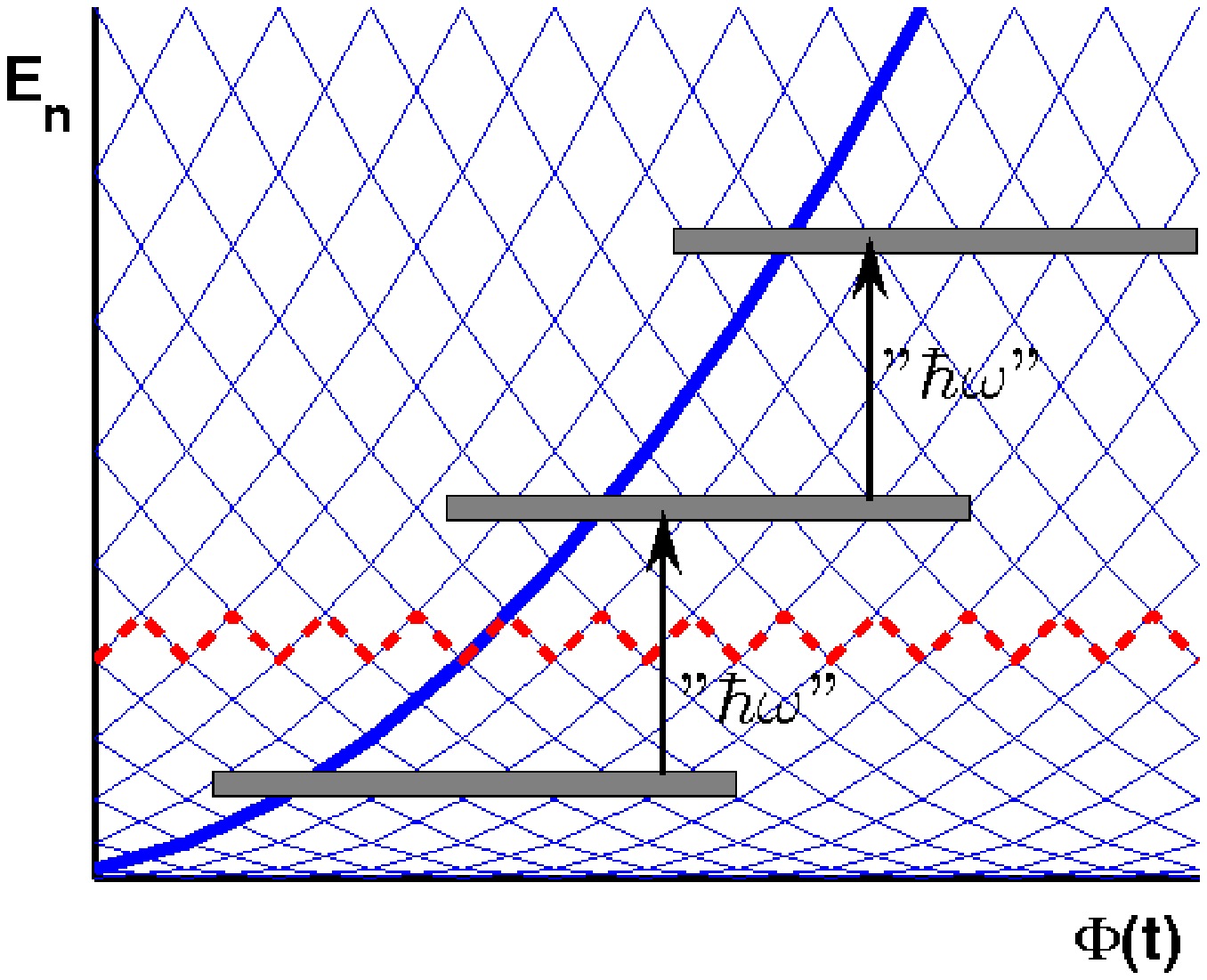}

{\footnotesize {\bf Fig.4:} 
(a) Left panel: The energy levels 
of a one dimensional box as a function of its width $L(t)$. 
For the purpose of comparison with other figures $L(t)$ is 
decreasing as we go to the right (the box becomes smaller) 
so the levels are going up.  The solid line 
illustrates adiabatic dynamics, while the "jumps" 
illustrate semiclassical dynamics. 
(b) Right panel: The energy levels of a one dimensional ring 
as a function of the Aharonov-Bohm flux. 
The solid line illustrates diabatic dynamics, 
while the "jumps" illustrate semiclassical dynamics. 
The dashed line illustrates adiabatic dynamics.}

\ \\ \ \\

\putgraph[width=0.8\hsize]{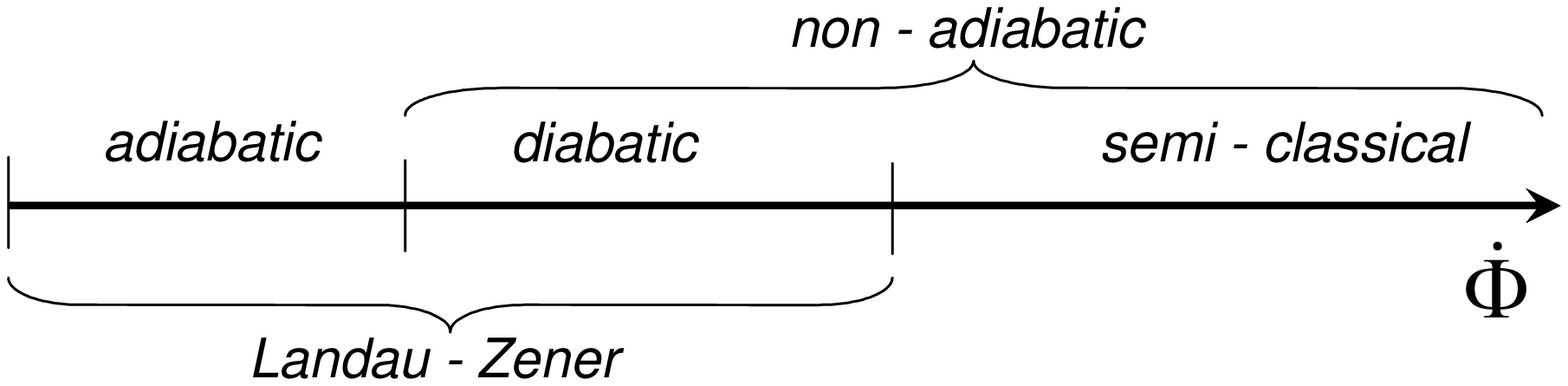}

{\footnotesize {\bf Fig.5:}
The three regimes in the EMF driven ring problem. See text.}

\ \\ \ \\

\putgraph[width=0.45\hsize]{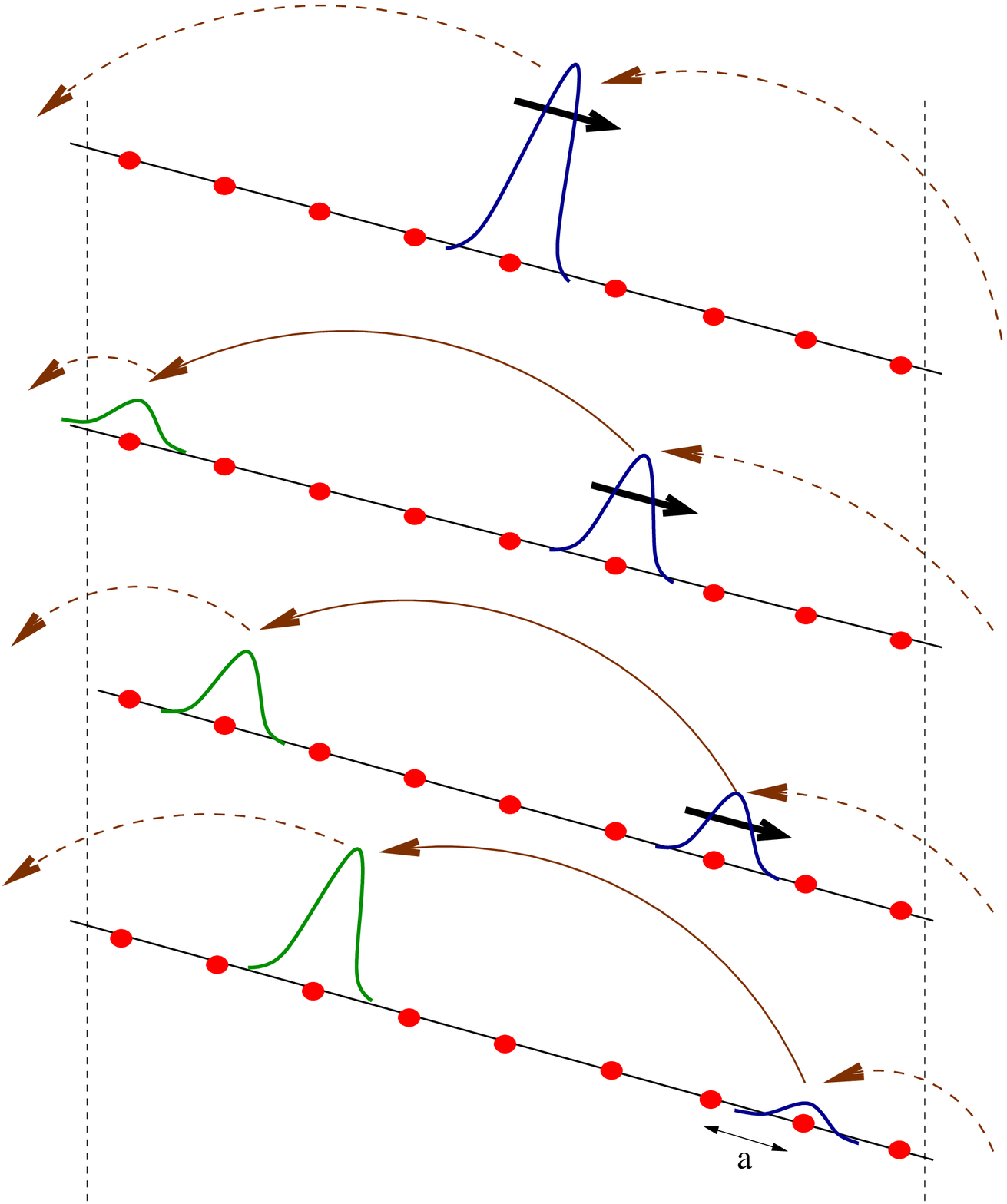}

{\footnotesize {\bf Fig.6:}
The unidirectional oscillations of  
Bloch electrons with $\propto 1/(n-m)$ hopping:  
As the wavepaket slides to the right 
it shrinks, while being re-injected on the left,  
where it re-emerges. This should be contrasted 
with conventional bi-directional oscillations 
of Bloch electrons with nearest neighbor hopping.}

}

\clearpage

\mpg{
\putgraph[width=0.45\hsize]{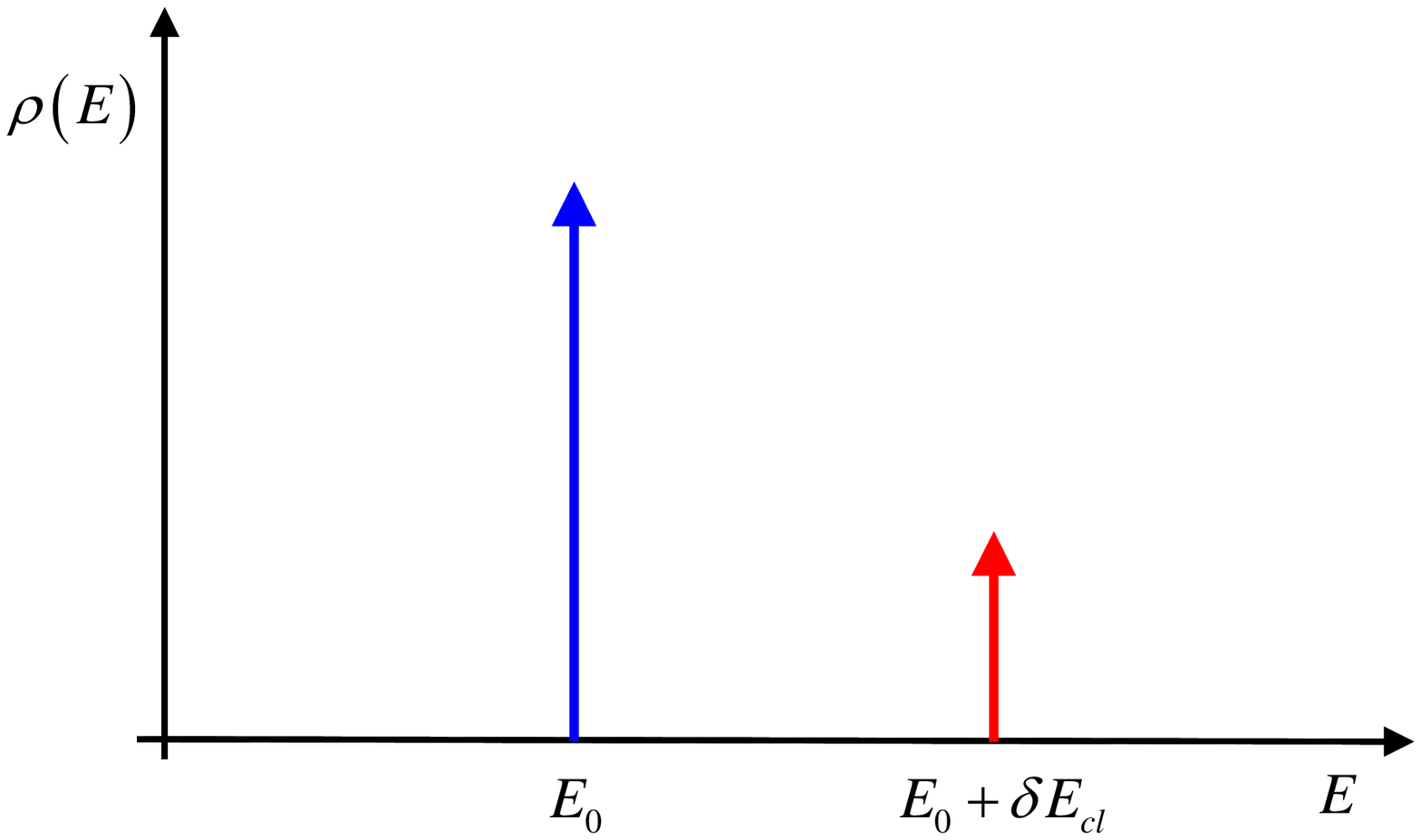}
\hfill
\putgraph[width=0.45\hsize]{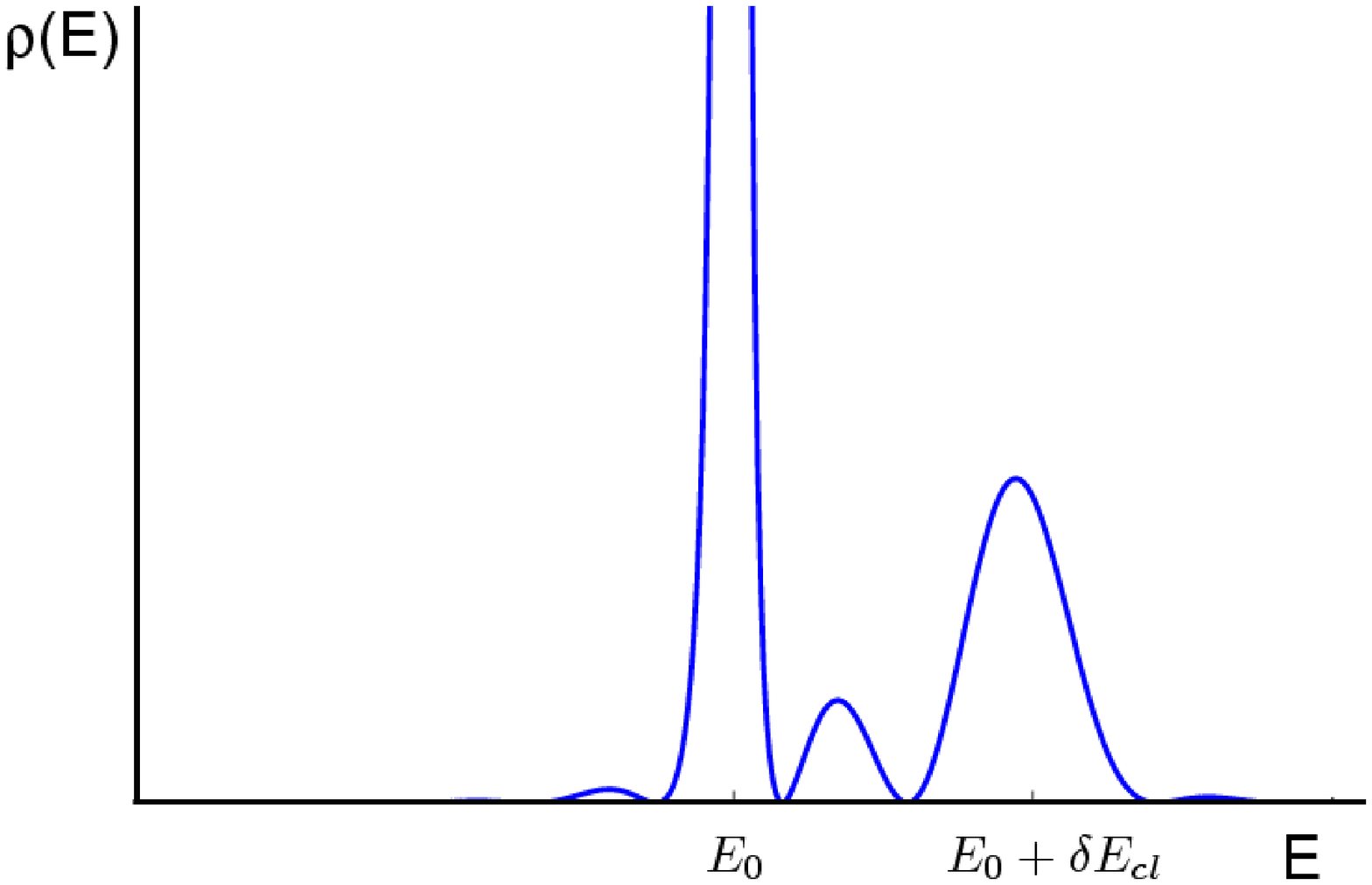}

{\footnotesize {\bf Fig.7:}
(a) Left panel: The classical energy distribution Eq.(\ref{e16})
some time after a step potential is turned on.
In this illustration $V_{\tbox{step}}{<}0$.  
(b) Right panel: the corresponding quantum mechanical  
energy distribution calculated with Eq.(\ref{e18}).}

}

\ \\

\mpg{
\putgraph[width=0.45\hsize]{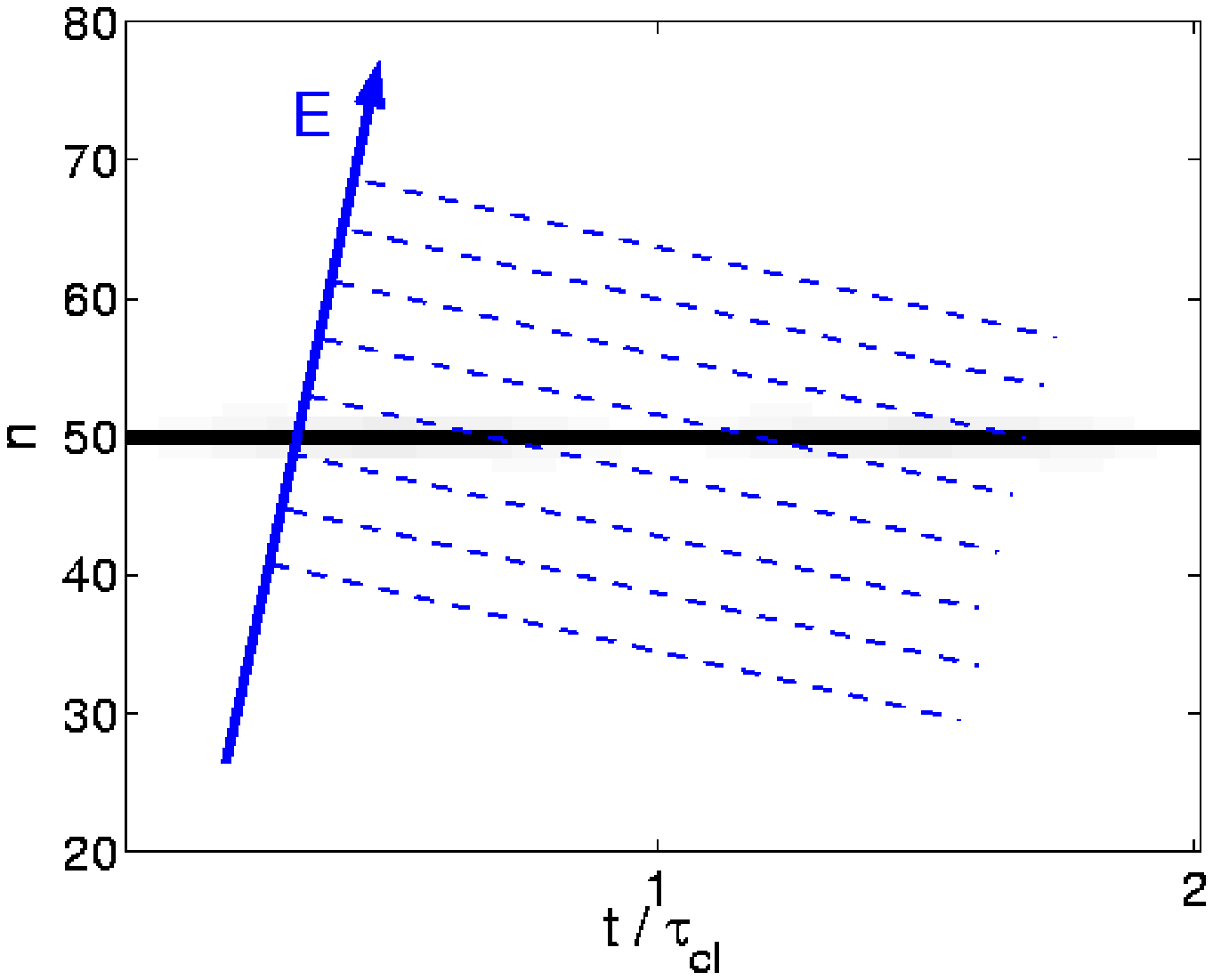}
\hfill
\putgraph[width=0.45\hsize]{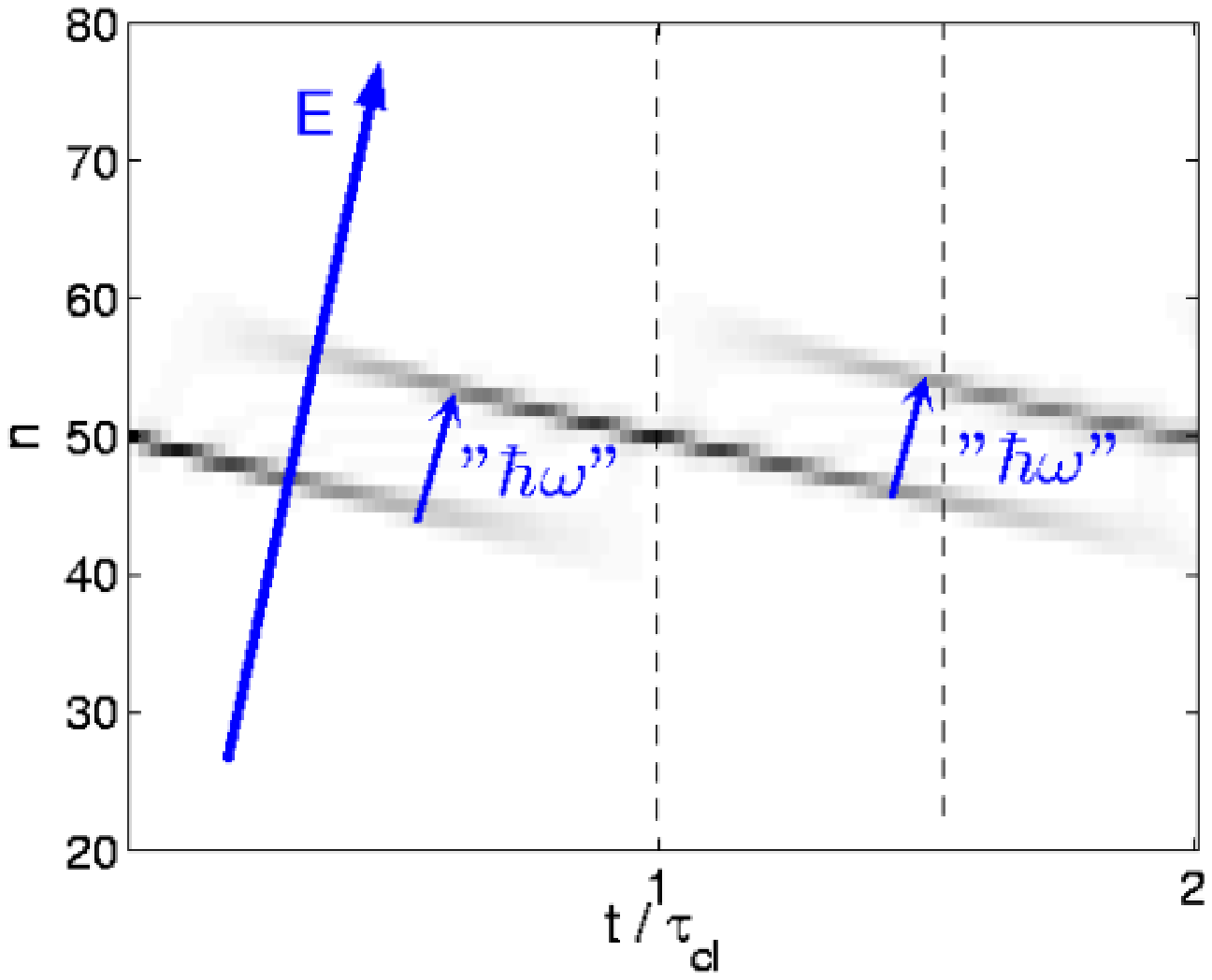}

{\footnotesize {\bf Fig.8:} 
Density plots of the probability distribution as a
function of time for the moving wall problem. 
{\bf (a)}~Left panel: Adiabatic regime. The probability stays 
at the same level. In order to clarify the connection 
with Fig.~4a we have added an~$E$ axis. 
The constant energy dashed lines are for guiding the eye.   
The populated adiabatic level goes up in energy which 
implies that the particle is steadily increasing its energy.   
{\bf (b)}~Right panel: semiclassical regime. 
The probability jumps in energy space. 
Note that with respect to the $E$~axis we 
have the steps of Fig.~4a. 
The parameters of these simulations  
were ${L {=} \mass {=} \hbar {=} 1}$ 
and $V_{\tbox{wall}} {=} 0.1\pi$ for~(a) 
and $V_{\tbox{wall}} {=} 5\pi$ for~(b).
Note that for $n {=} 50$ the classical period 
is $\tau_{\tbox{cl}} {=} 0.0127$. The vertical 
dashed lines indicate two representative 
times $t {=} \tau_{\tbox{cl}}$ 
and $t {=} 1.5\tau_{\tbox{cl}}$. } 

}

\ \\

\mpg{
\putgraph[width=0.45\hsize]{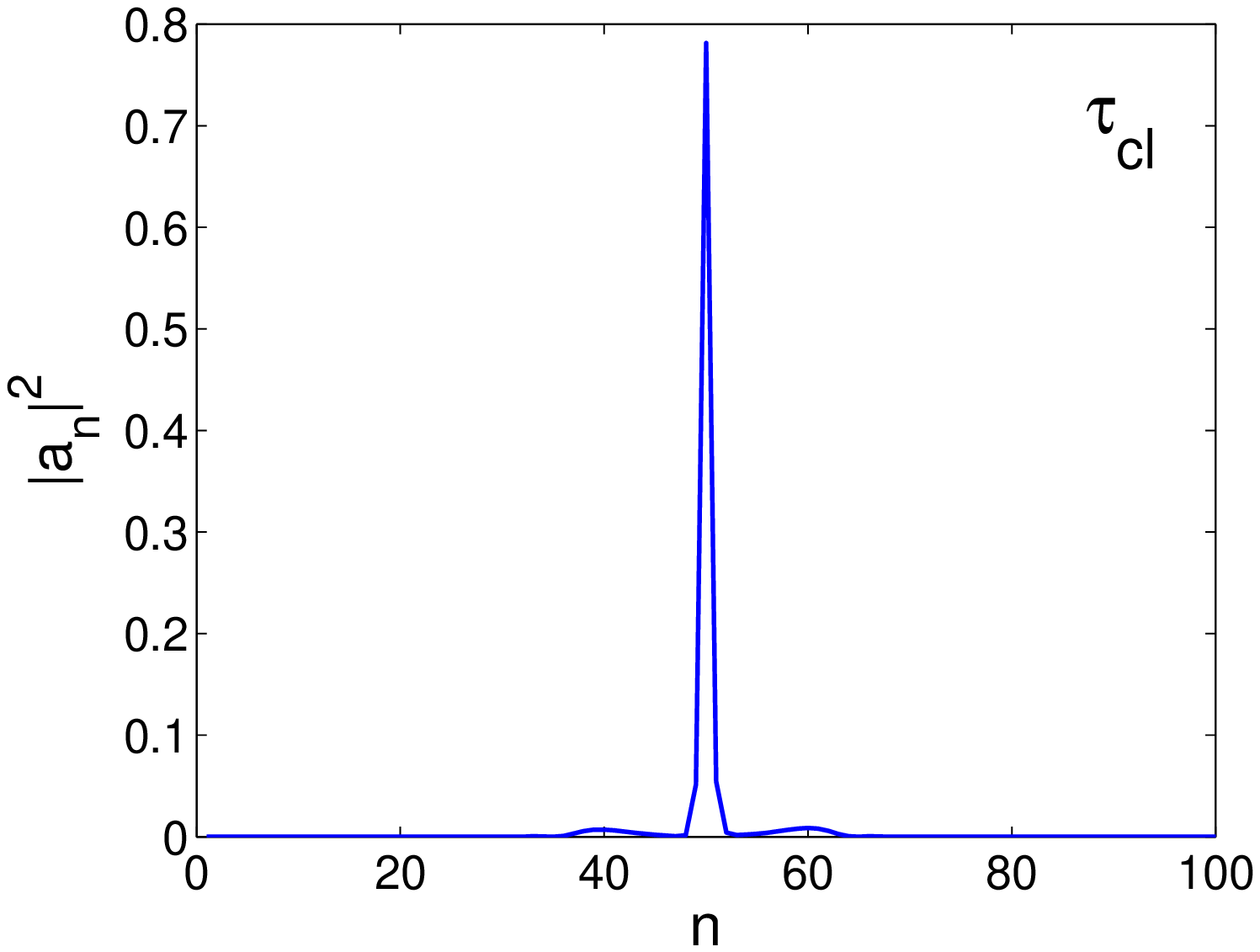}
\hfill
\putgraph[width=0.45\hsize]{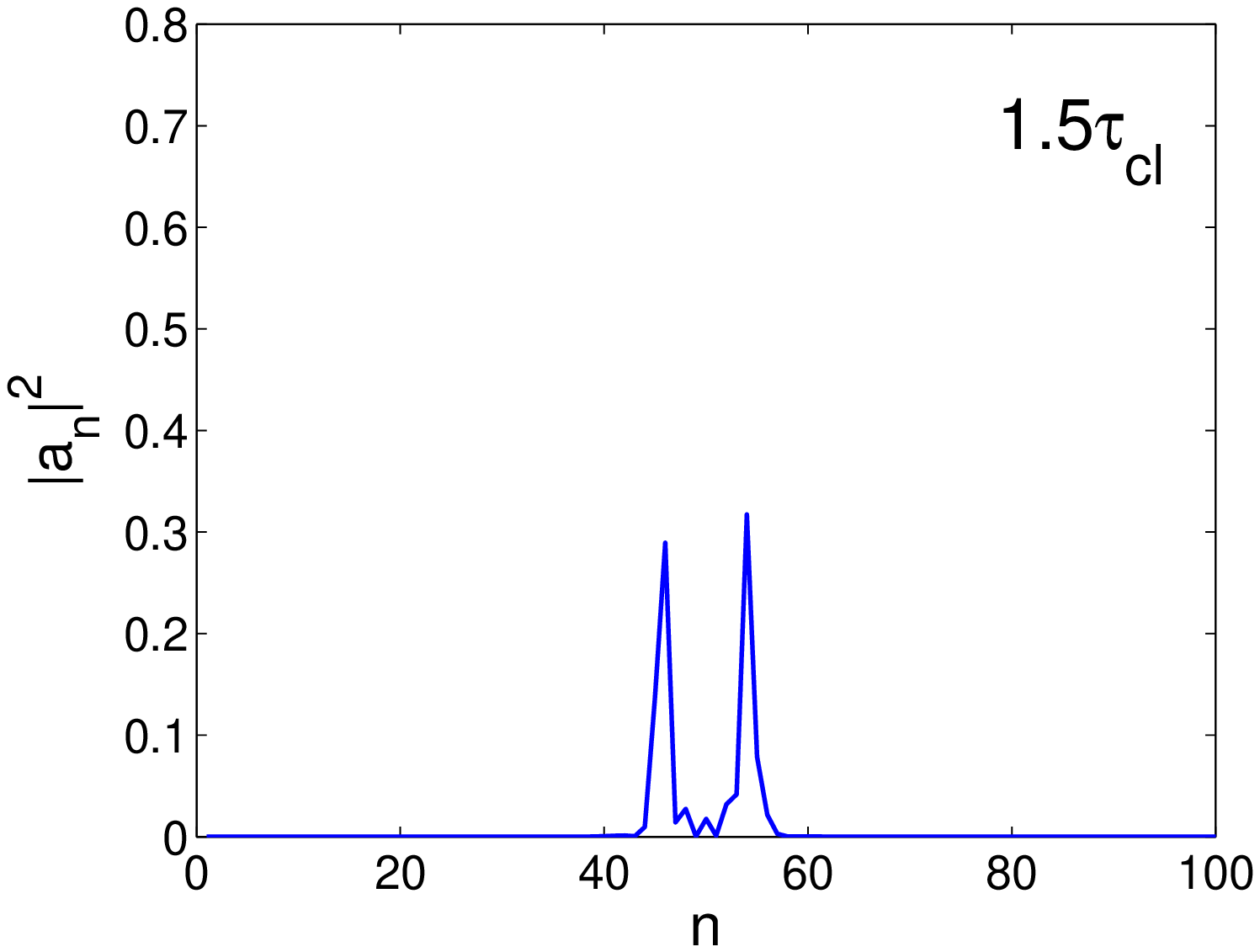}

{\footnotesize {\bf Fig.9:} 
Plots of the probability distribution $|a_n|^2$ 
at the two representative times that were indicated 
by the vertical dashed lines in the previous figure.
In the classical limit the energy distribution 
consists of delta peaks instead of broadened peaks,  
in complete analogy with Fig.~7.}

}

\ \\ \ \\

\mpg{
\noindent
\putgraph[width=0.45\hsize]{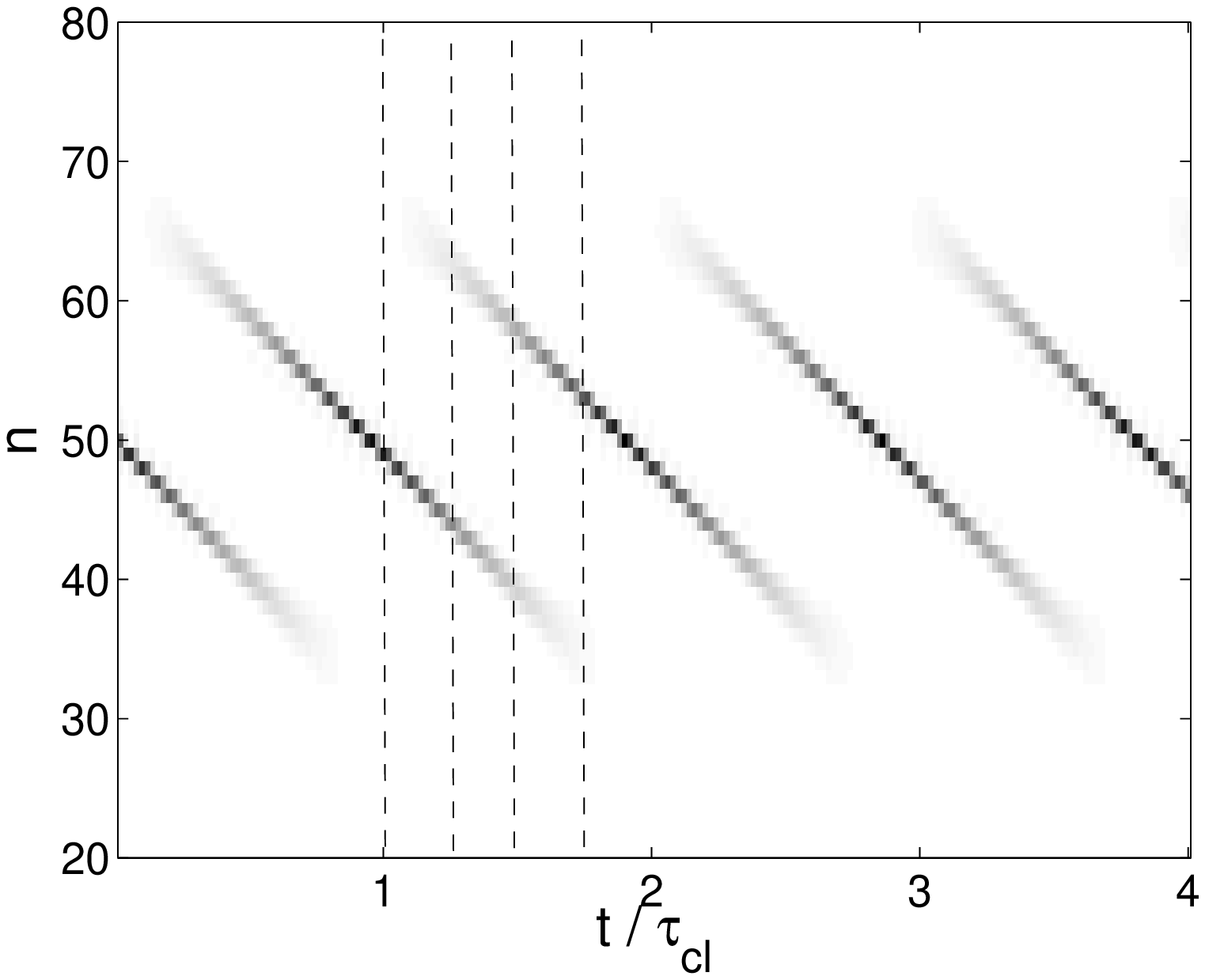}
\hfill
\putgraph[width=0.45\hsize]{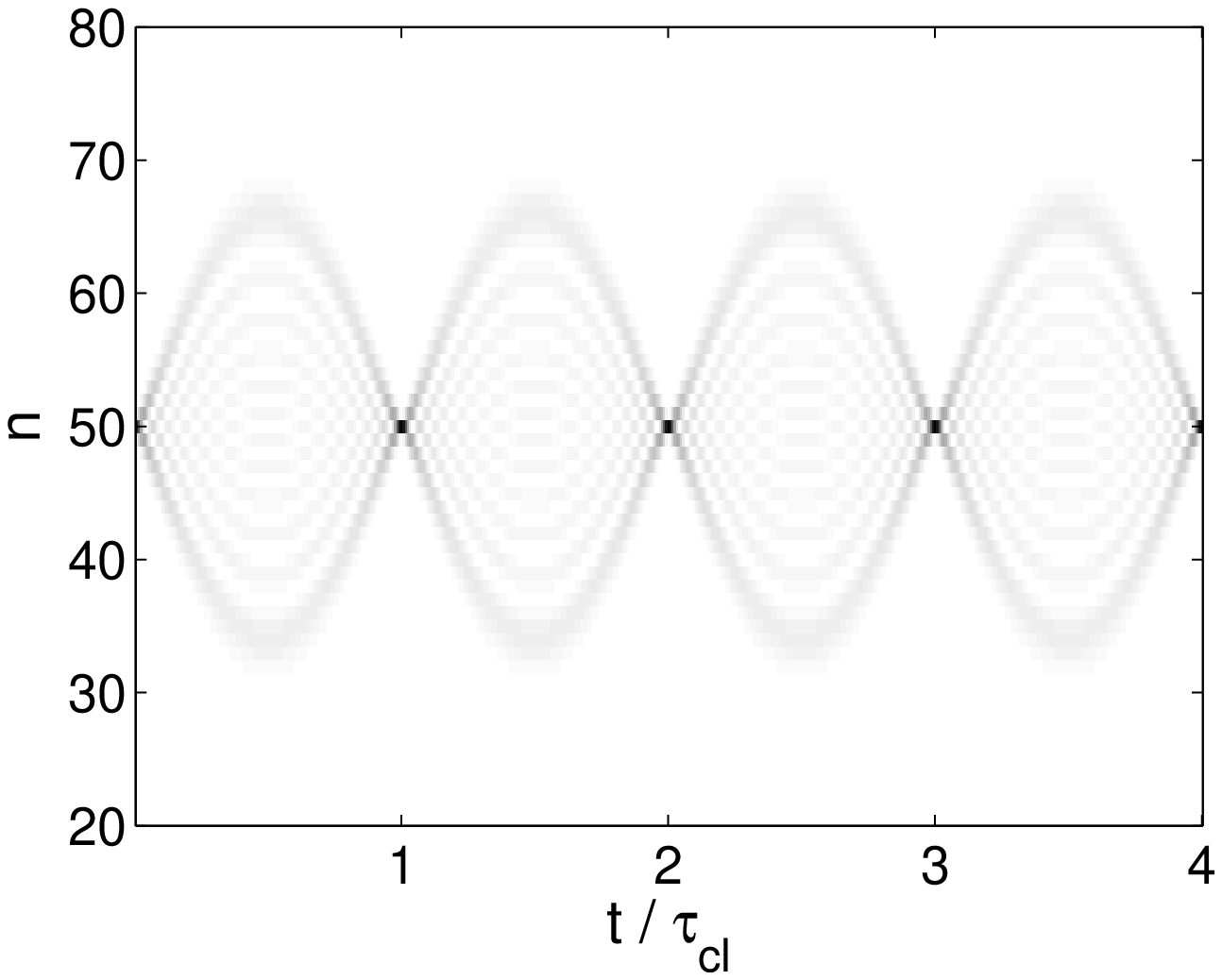}

{\footnotesize {\bf Fig.10:} 
(a) Density plots of the probability distribution 
as a function of time for the EMF driven ring problem.
See the caption of Fig~8 for presentation details.
The parameters are $L = \mass = \hbar = e = 1$
with $V_{\tbox{EMF}}=588840$. 
Note that for $n=50$ the classical period 
is $\tau_{\tbox{cl}}=0.00021$. 
This is approximately the same as solving the wavepacket 
dynamics for Bloch electron in electric field   
Eq.(\ref{e24}) with $\alpha=93717$ and $\varepsilon=31416$.
(b)~Wavepacket dynamics for Bloch electron in electric 
field with near-neighbor hopping.
The parameters are $\mass =\hbar = e = 1$ 
with $\alpha=14139$ and $\varepsilon=3141.6$.}

}

\ \\ \ \\

\mpg{
\noindent \putgraph[width=0.45\hsize]{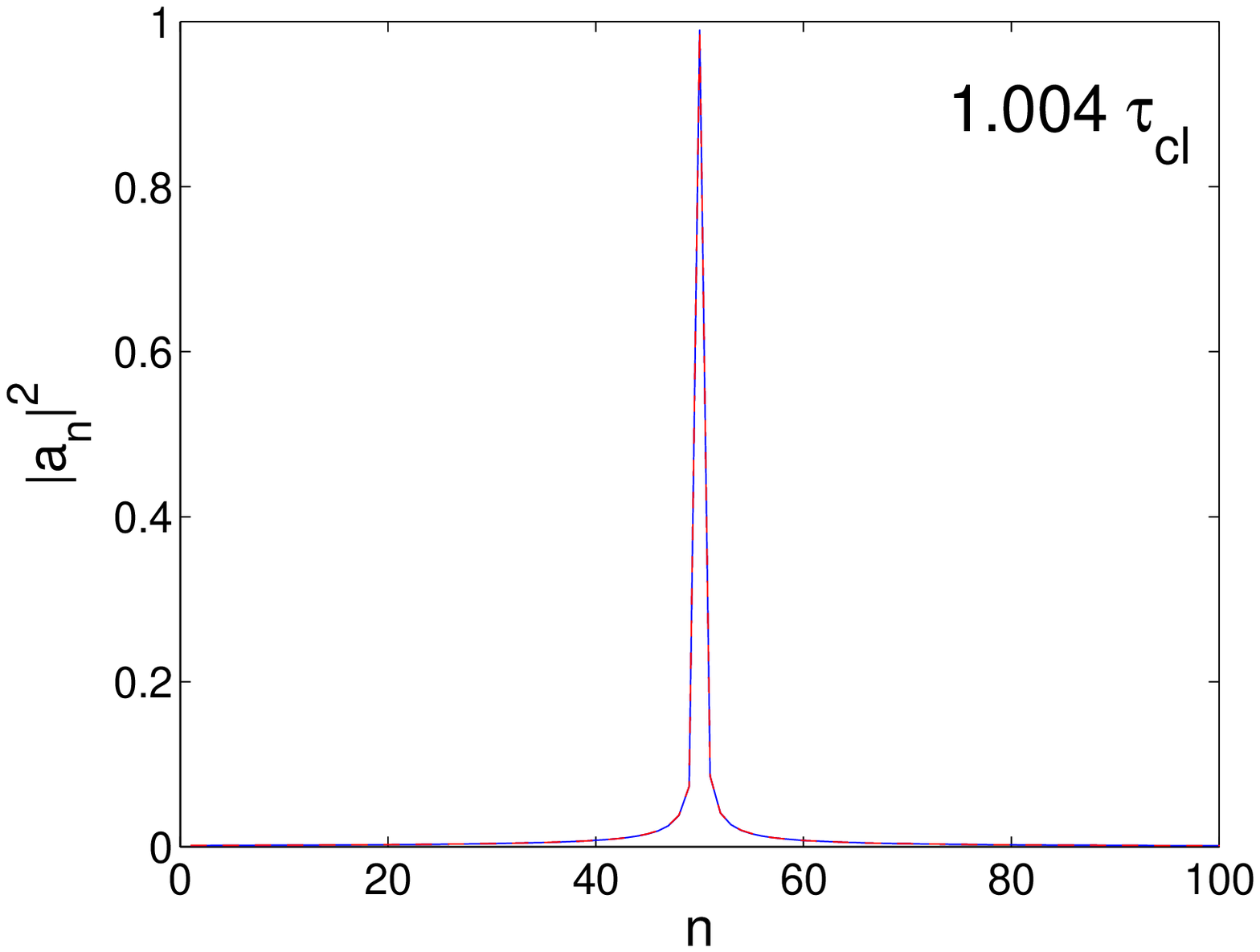} \hfill
\putgraph[width=0.45\hsize]{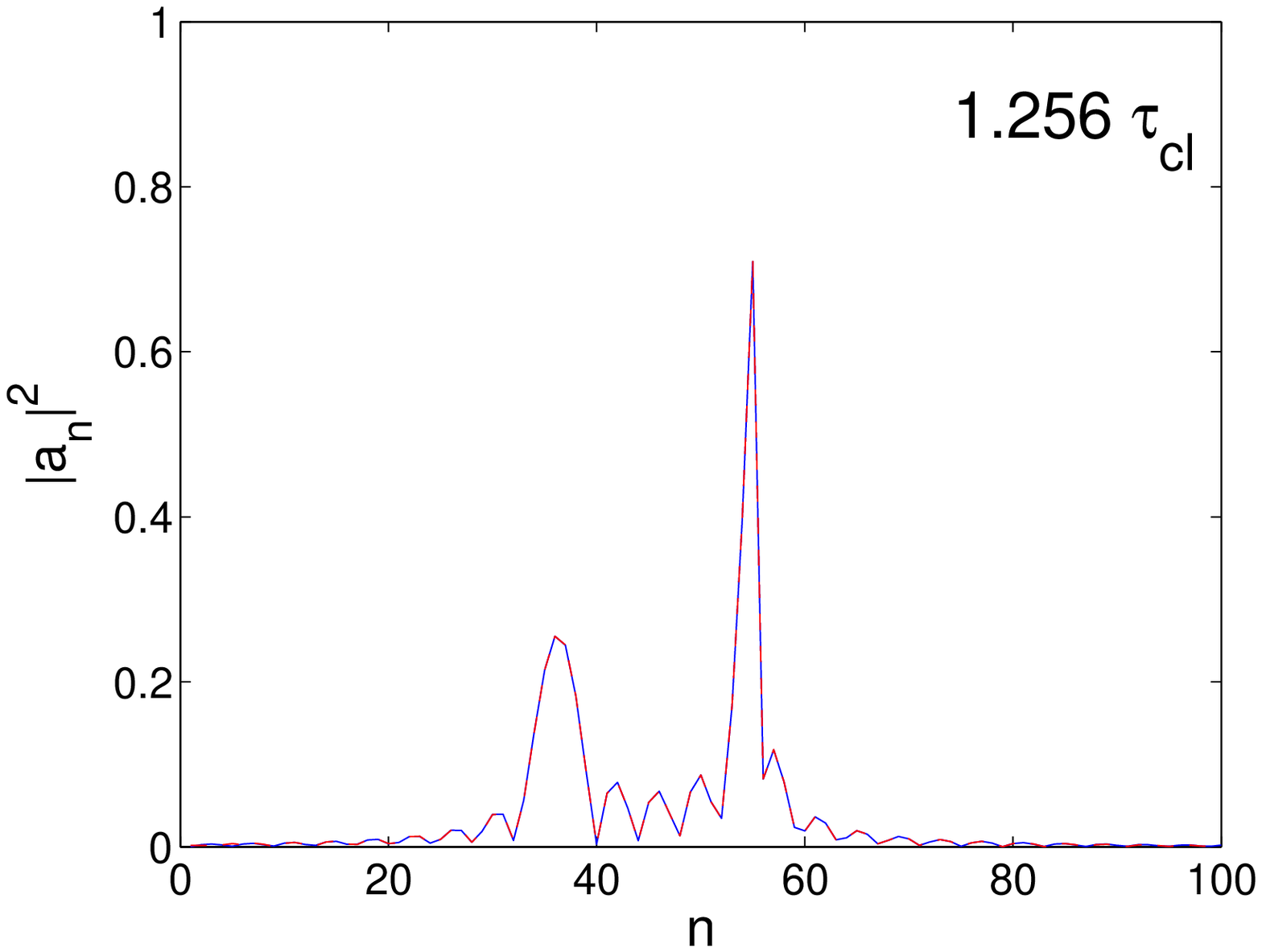} \hfill
\putgraph[width=0.45\hsize]{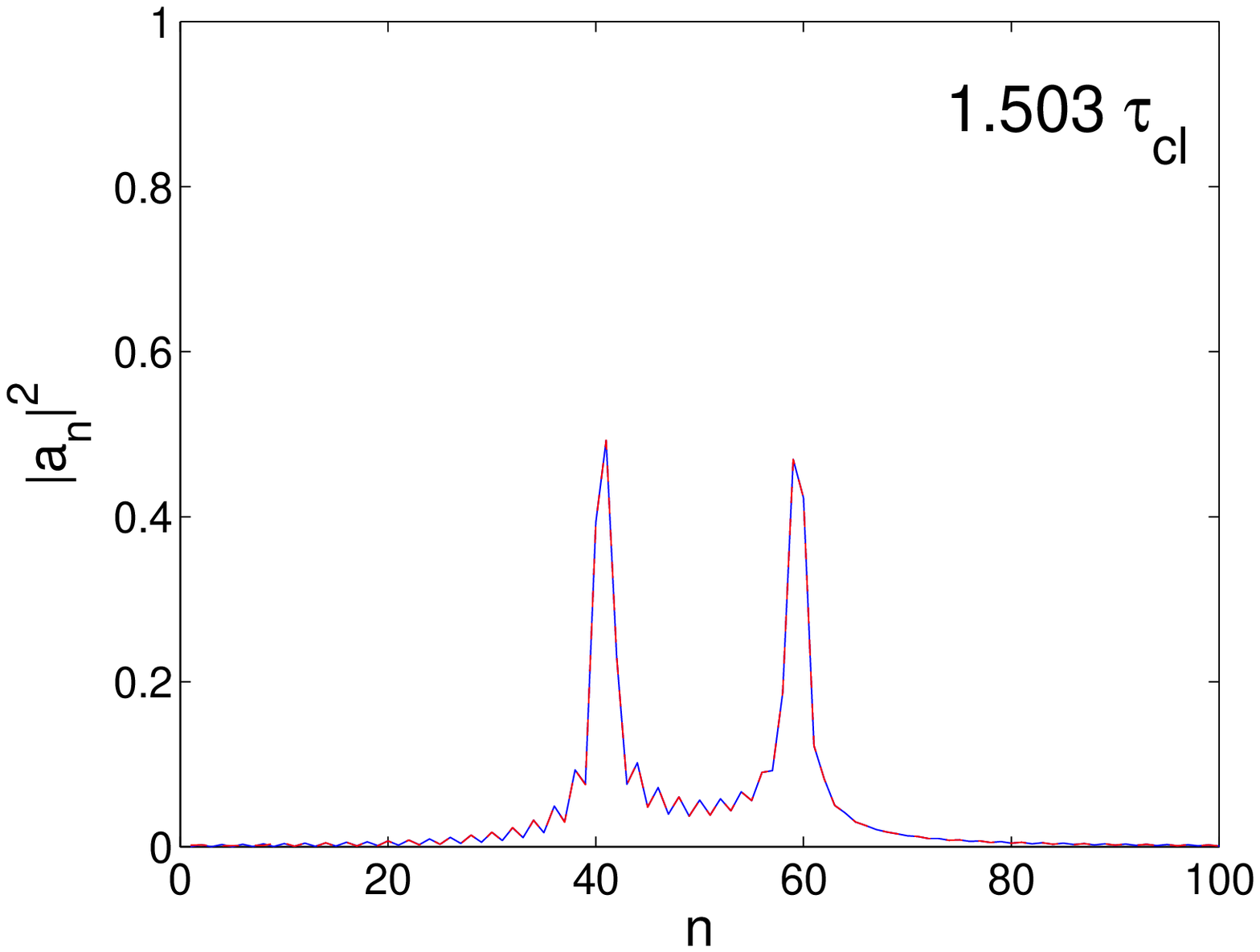} \hfill
\putgraph[width=0.45\hsize]{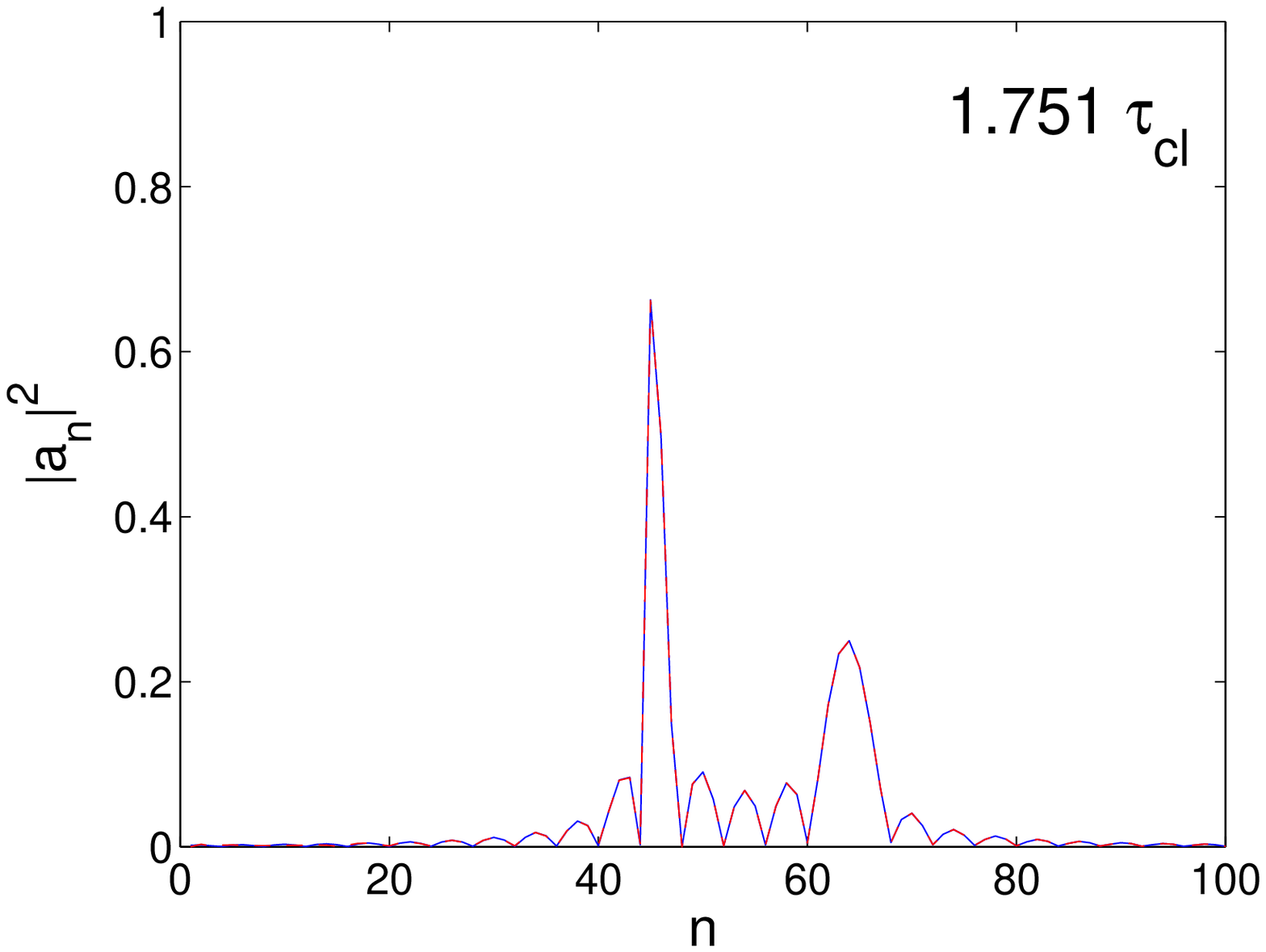}

{\footnotesize {\bf Fig.11:} 
Plots of the probability distribution $|a_n|^2$ 
at representative times (as indicated). 
The solid lines are the solution of Eq.(\ref{e24}) 
for Bloch electrons, while the dotted lines 
are the exact numerical solutions for 
the EMF driven ring, taking into account 
the quadratic (rather than linear) dependence 
of the eigen-energies on $\Phi$. }

}

\end{document}